\documentclass[%
reprint,superscriptaddress,
amsmath,amssymb,aps,
]{revtex4-1}

\usepackage{graphicx}
\usepackage{dcolumn}
\usepackage{bm}
\usepackage{float}
\usepackage{xcolor}

\begin{document}

\preprint{APS/123-QED}

\title{Iso-entropy partially coherent optical fields that cannot be inter-converted unitarily}

\author{Mitchell Harling}
\affiliation{PROBE Lab, School of Engineering, Brown University, Providence, RI 02912, USA}%
\author{Varun A. Kelkar}
\affiliation{Department of Electrical and Computer Engineering, University of Illinois at Urbana-Champaign, Urbana, IL 61801, USA}%
\affiliation{Currently at Algorithmic Systems Group, Analog Garage, Analog Devices, Inc., Boston, MA 02110, USA}%
\author{Kimani C. Toussaint, Jr.}
\affiliation{PROBE Lab, School of Engineering, Brown University, Providence, RI 02912, USA}%
\affiliation{Brown-Lifespan Center for Digital Health, Providence, RI 02912, USA}%
\author{Ayman F. Abouraddy}%
\affiliation{CREOL, The College of Optics \& Photonics, University of Central Florida, Orlando, FL 32816, USA}%

\date{\today}

\begin{abstract}
For partially coherent optical fields in which a single binary degree of freedom (DoF) is relevant, such as polarization, entropy uniquely identifies the class of optical fields that can be converted into each other via unitary transformations. However, when multiple DoFs are taken into consideration, entropy no longer serves this purpose. We investigate the structure of the family of iso-entropy partially coherent optical fields defined by two binary DoFs (polarization and two spatial modes) and described by a $4\times4$ coherence matrix $\mathbf{G}$. We find that the rank of $\mathbf{G}$ (the number of its non-zero eigenvalues) plays a critical role in this context: whereby any pair of iso-entropy rank-2 fields can be converted into each other unitarily, this is not necessarily the case for a pair of rank-3 or rank-4 fields. Furthermore, unitary transformations between iso-entropy fields of different ranks are strictly forbidden. Instead, such conversions require entropy-maintaining non-unitary transformations that potentially combine filtering projections and randomizing operations. We experimentally synthesize partially coherent iso-entropy optical fields of all ranks, and tomographically reconstruct their coherence matrices. Moreover, we steer the coherence matrix over iso-entropy trajectories that maintain a fixed rank (intra-rank conversion) or that involve changes in the rank (inter-rank conversion). These findings offer a new perspective for the potential utility of partially coherent light in optical communications and sensing. 
\end{abstract}

\maketitle

\section{Introduction}

Statistical fluctuations underpin the partial coherence of optical fields \cite{Goodman15Book}. The study of optical coherence over the past century has provided a comprehensive framework for describing statistical optical phenomena \cite{Zernike38P,Mandel65RMP,Wolf07Book} -- from interference and laser speckle \cite{goodman1976some,Goodman07Book,Saleh07Book} to higher-order statistical effects \cite{Glauber63PR,Bromberg14PRL,Kondakci15NP,Kondakci16Optica,Kondakci17SR,Wang19PRL,Podivilov22PRL,Han23PRL}. Traditionally, the coherence of each degree of freedom (DoF) of the optical field has been investigated separately, whether for the spatial, spectral/temporal, or polarization DoF. However, it is now becoming clear that taking multiple DoFs of the optical field jointly into consideration opens up new vistas for studying optical coherence \cite{Gori98OL,Wolf03PLA,Refregier05OE,Gori06OL,Gori07OL,Luis07OL,Kagalwala13NP,Abouraddy17OE}, and is making possible a variety of new applications \cite{Qian11OL,Toppel14NJP,Aiello2015cc,Otte18LSA,Mamani18JB,kondakci2019classical,Toninelli19AOP,YaoLi20APL,Shen21LSA,Shen21PRR,Hall22JOSAA,Aiello22NJP}.

For concreteness, consider a field characterized by two binary DoFs (polarization and two spatial modes), and is thus represented by a $4\times4$ coherence matrix $\mathbf{G}$ \cite{Kagalwala13NP}. We have found recently that the \textit{rank} of this coherence matrix -- the number of its non-zero eigenvalues (a parameter not investigated hitherto) -- helps identify novel features of the field \cite{Harling24PRA}. In the scenario studied here of two binary DoFs, the coherence rank can take on values 1, 2, 3, or 4, thus leading to a fourfold taxonomy of optical fields relative to their coherence rank. Partially coherent optical fields can be distinguished by structural features that depend on their coherence rank. For example, a novel insight uncovered in \cite{Harling24PRA} is that rank-2 fields are \textit{always} separable with respect to the two DoFs, whereas rank-3 fields are \textit{never} separable. Another salutary property of the coherence rank is that it is a unitary invariant of the field; i.e., the rank is invariant under deterministic unitary transformations (reversible energy-preserving transformations, henceforth `unitaries' for brevity) that modify one DoF or the other, that modify both DoFs independently, or that couple them in their joint space. Incidentally, the coherence rank of each DoF separately from the other (determined from the associated `reduced' coherence matrix after tracing out the other DoF) is \textit{not} invariant under some of these unitaries -- particularly unitaries that couple the two DoFs \cite{Kagalwala13NP,Abouraddy17OE,Okoro17Optica}. 

Another critical descriptor of a partially coherent field comprising multiple DoFs -- besides its coherence rank -- is its entropy $S$, which is also a unitary invariant of the field, and is taken to quantify the field fluctuations \cite{Brosseau06PO, refregier2008irreversible}. The coherence rank has profound implications for the range of possible reversible conversion of entropy between DoFs via unitaries \cite{Okoro17Optica,harling2022reversible}. We have shown in \cite{Harling24PRA} that the entropy of a rank-2 field that is initially shared between the two DoFs -- no matter how high -- can \textit{always} be reversibly concentrated into one DoF, leaving the other DoF free of statistical fluctuations. In contrast, the entropy of a rank-3 field -- no matter how low -- can\textit{not} be concentrated into a single DoF, giving rise to what we have denoted as `locked' entropy \cite{Harling24PRA}.

In the case of a single binary DoF, such as polarization \cite{Brosseau06PO} or a pair of spatial modes \cite{Eberly17Optica,DeZela18Optica,Abouraddy19Optica,Halder21OL}, the entropy can be used to uniquely identify an equivalence class of optical fields. In other words, any two optical fields in this scenario endowed with the same entropy (henceforth `iso-entropy' fields) can \textit{always} be converted into each other via a unitary. 
The question we pose here in the context of \textit{two} binary DoFs is the following: does the entropy $S$ remain a unique descriptor for the equivalence class of \textit{all} optical fields? In other words, can iso-entropy optical field configurations always be converted into each other via unitaries? Or, do there exist iso-entropy fields that cannot be inter-converted into each other except by resorting to entropy-preserving \textit{non}-unitary transformations?

Here, we show that the entropy $S$ of a partially coherent optical field encompassing multiple DoFs does \textit{not} delineate the class of fields that can be inter-converted unitarily. Therefore, there do indeed exist iso-entropy partially coherent field configurations that cannot be converted into each other except by entropy-preserving non-unitary transformations, which typically comprise randomizing transformations that increase the entropy and projective filters that decrease it. 

We find that the coherence rank plays a key role in this regard. We distinguish between two scenarios: `intra-rank' transformations (involving iso-entropy fields of the \textit{same} rank), and `inter-rank' transformations (involving iso-entropy fields of \textit{different} ranks). In the latter case, it is never possible for two iso-entropy fields of different ranks to be transformed into each other via unitaries. In other words, unitary inter-rank transformations are forbidden, and non-unitary transformations are required instead. Moreover, transforming between lower-rank to higher-rank iso-entropy fields requires a combination of randomizing (entropy-increasing) and filtering (entropy-decreasing) systems, whereas only an entropy-preserving filtering system is needed for going in the opposite direction (from higher-rank to lower-rank iso-entropy fields). With regard to \textit{intra}-rank transformations, the possibility of relying solely on unitaries to convert iso-entropy fields into each other depends critically on the coherence rank. Trivially, all rank-1 fields (fully coherent fields) can be converted into each other via unitaries. The same applies to all rank-2 optical fields, where the entropy is a unique identifier of the field structure (just as in the case of a single binary DoF).
However, this is not the case for rank-3 and rank-4 fields. Iso-entropy rank-3 fields can be assembled into a one-parameter 
family of fields, where fields associated with different values of this parameter cannot be converted into each other unitarily. Iso-entropy rank-4 fields are assembled into a two-parameter family of fields, where fields associated with different values of the two parameters cannot be converted into each other unitarily. 

We have validated these theoretical results experimentally by synthesizing partially coherent optical fields of different rank and entropy (a total of 114 distinct partially coherent field configurations), and reconstructing their $4\!\times\!4$ coherence matrix $\mathbf{G}$ via optical coherency matrix tomography (OCmT) \cite{Abouraddy14OL,Kagalwala15SR}. The results are depicted in a three-dimensional (3D) geometric space spanned by three eigenvalues of the trace-normalized $\mathbf{G}$. In addition,  using deterministic entropy-preserving non-unitary transformations, we have steered rank-3 fields across the one-parameter curve of iso-entropy intra-rank fields, and rank-4 fields across the two-parameter iso-entropy intra-rank surface. Finally, we have steered the coherence matrix across an inter-rank iso-entropy trajectory. Starting from a rank-4 field, we produce a rank-3 field, from which we then produce a rank-2 field -- all having the same entropy. These inter-rank transformations utilize only deterministic non-unitary transformations. We then reverse the process, and starting with a rank-2 field we produce a rank-3 field, from which we then produce a rank-4 field -- all having once again the same entropy. In this case, however, randomizing transformations are required. Before concluding, we explore the implications of these findings for using partially coherent multi-DoF fields in optical communications.

\section{Geometric representation of iso-entropy fields}

\subsection{Vector-space formulation and entropy of partially coherent optical fields}

We consider optical fields characterized by two binary DoFs, taken here to be polarization (spanned by horizontal, H, and vertical, V, polarization components) and two spatial modes (labeled `$a$' and `$b$'). The first-order coherence for this field is described by a $4\!\times\!4$ unity-trace coherence matrix $\mathbf{G}$  that is Hermitian and positive semi-definite \cite{Gori06OL,Kagalwala13NP,Abouraddy17OE}:
\begin{equation}
\mathbf{G}=\left(\begin{array}{cccc}
G_{aa}^{\mathrm{HH}}&G_{aa}^{\mathrm{HV}}&G_{ab}^{\mathrm{HH}}&G_{ab}^{\mathrm{HV}}\vspace{0.1cm}\\
G_{aa}^{\mathrm{VH}}&G_{aa}^{\mathrm{VV}}&G_{ab}^{\mathrm{VH}}&G_{ab}^{\mathrm{VV}}\vspace{0.1cm}\\
G_{ba}^{\mathrm{HH}}&G_{ba}^{\mathrm{HV}}&G_{bb}^{\mathrm{HH}}&G_{bb}^{\mathrm{HV}}\vspace{0.1cm}\\
G_{ba}^{\mathrm{VH}}&G_{ba}^{\mathrm{VV}}&G_{bb}^{\mathrm{VH}}&G_{bb}^{\mathrm{VV}}
\end{array}\right),
\end{equation}
where $G_{kl}^{ij}\!=\!\langle E_{k}^{i}(E_{l}^{j})^{*}\rangle$, $\langle\cdot\rangle$ denotes an ensemble average, $i,j\!=\!\mathrm{H},\mathrm{V}$, and $k,l\!=\!a,b$. We define the entropy for $\mathbf{G}$ as
\begin{equation}\label{eq:EntropyDefinition}
S=-\sum_{i=1}^{4}\lambda_{i}\mathrm{log}_{2}\lambda_{i},
\end{equation}
where $\{\lambda_{1},\lambda_{2},\lambda_{3},\lambda_{4}\}$ are the eigenvalues of $\mathbf{G}$, and $\sum_{i=1}^{4}\lambda_{i}\!=\!1$. In general, the entropy for two binary DoFs lies in the range $0\!\leq\!S\!\leq\!2$~bits. The condition $S\!=\!0$ indicates the complete absence of statistical fluctuations in the field (coherent fields), whereas $S\!=\!2$~bits corresponds to maximal statistical fluctuations across both DoFs (incoherent fields). Of course, each DoF separately can carry at most 1~bit of entropy. This formulation is the foundation for our previous work on the reversible exchange of entropy between the DoFs of the field \cite{Okoro17Optica, harling2022reversible, Harling23JO, Harling24PRA}.

Any two fields whose coherence matrices $\mathbf{G}_{1}$ and $\mathbf{G}_{2}$ can be inter-converted via a similarity transformation $\mathbf{G}_{2}\!=\!\hat{U}_{12}\mathbf{G}_{1}\hat{U}_{12}^{\dag}$, where $\hat{U}_{12}$ is a unitary transformation, will have the same entropy because the eigenvalues are invariant under such a transformation \cite{Abouraddy17OE}. We refer to the two optical fields represented by $\mathbf{G}_{1}$ and $\mathbf{G}_{2}$ as `iso-entropy' fields. Of course, the structure of the matrices $\mathbf{G}_{1}$ and $\mathbf{G}_{2}$ may differ significantly, and they would thus represent very different field configurations. Nevertheless, they share the same eigenvalues, and are thus endowed with the same entropy. 

\subsection{Geometric representation of iso-entropy fields}

To emphasize the characteristics of the entropy, we introduce a \textit{geometric} representation for the coherence matrix that relies solely on its eigenvalues. Consider \textit{diagonal} coherence matrices of the form:
\begin{equation}\label{eq:Diagonal}
\mathbf{G}=\left(\begin{array}{cccc}
\lambda_{1}&0&0&0\\
0&\lambda_{2}&0&0\\
0&0&\lambda_{3}&0\\
0&0&0&\lambda_{4}
\end{array}\right)=\mathrm{diag}\{\lambda_{1},\lambda_{2},\lambda_{3},\lambda_{4}\},
\end{equation}
where $0\!\leq\!\lambda_{j}\!\leq\!1$ and $\sum_{j=1}^{4}\lambda_{j}\!=\!1$. This diagonal coherence matrix represents an entire `class' of coherence matrices that can all be converted into each other via unitaries. We take the diagonal coherence matrix in Eq.~\ref{eq:Diagonal} to be representative of this entire equivalence class, and thus concern ourselves henceforth only with such diagonal matrices. We define the \textit{rank} of $\mathbf{G}$ as the number of its non-zero eigenvalues, which can thus take on the values 1, 2, 3, or 4, denoted rank-1, rank-2, rank-3, or rank-4, respectively.

\begin{figure}[t!]
    \includegraphics[width=8.6cm]{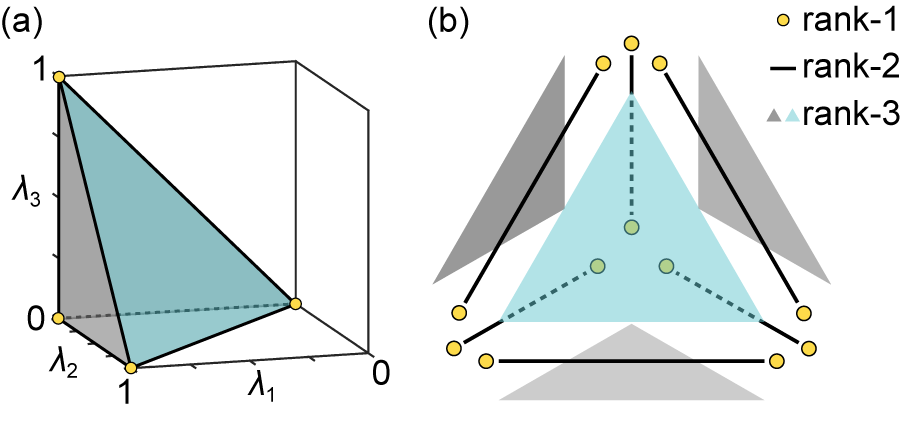}
    \caption{(a) The geometric domain (in the form of a triangular pyramid) corresponding to all $4\times4$ coherence matrices $\mathbf{G}$, restricted to a representation in terms of three of its eigenvalues $0\!<\!\lambda_{1},\lambda_{2},\lambda_{3}\!<\!1$, with the fourth eigenvalue given by $\lambda_{4}\!=\!1-(\lambda_{1}+\lambda_{2}+\lambda_{3})$. In this representation, a coherence matrix $\mathbf{G}$ corresponds to a point in this geometric domain. (b) The faces, edges, and vertices of the triangular pyramid in (a) have been translated from their original positions for visualization. The front face is an equilateral triangle (shown in slate blue), and the other faces are right-angled isosceles triangles (shown in gray). The coherence matrices for rank-1 fields correspond to the four vertices; rank-2 to the six edges; rank-3 to the four faces; and rank-4 to the volume within the triangular pyramid where $\lambda_{1}+\lambda_{2}+\lambda_{3}\!<\!1$.}
    \label{fig1}
\end{figure}

Consider a four-dimensional (4D) space spanned by the parameters $\{\lambda_{1},\lambda_{2},\lambda_{3},\lambda_{4}\}$. Each coherence matrix $\mathbf{G}$ corresponds to a point in this space, and the constraint $\sum_{j=1}^{4}\lambda_{j}\!=\!1$ entails that coherence matrices are restricted to a subspace in the form of a hyperplane. Because it is difficult to visualize this geometric structure in 4D, we restrict ourselves to a 3D space spanned by only $\{\lambda_{1},\lambda_{2},\lambda_{3}\}$, and rely on the restriction $\lambda_{4}\!=\!1-\sum_{j=1}^{3}\lambda_{j}$. The hyperplane in the full 4D space is projected in the restricted 3D space onto the volumetric structure shown in Fig.~\ref{fig1}(a): a triangular pyramid in which three faces are right-angled isosceles triangles, and the fourth face is an equilateral triangle. Each point in this volume corresponds to a diagonal coherence matrix that represents a class of fields that share the same eigenvalues (and thus in turn the same rank and entropy). Note however that coherence matrices after permutations of the eigenvalues are not represented by the same point. For example, $\mathbf{G}\!=\!\mathrm{diag}\{1,0,0,0\}$ and $\mathbf{G}\!=\!\mathrm{diag}\{0,1,0,0\}$ correspond in the structure shown in Fig.~\ref{fig1}(a) to two different vertices (1, 0, 0) and (0, 1, 0), respectively.

This geometric structure is instructive because fields of different rank correspond to distinct geometric features of this representation. We explode the pyramid structure in Fig.~\ref{fig1}(a) in terms of its vertices, edges, and faces, as shown in Fig.~\ref{fig1}(b); the volume enclosed in the structure is omitted for clarity. Each geometric feature isolated in Fig.~\ref{fig1}(b) corresponds to differently ranked optical fields. 
Furthermore, the structure enables visualization of the entropy as a function of the eigenvalues, which is illustrated on the faces in Fig \ref{fig2}(a) and Fig. \ref{fig2}(b), and for iso-entropy surfaces within the volume in Fig \ref{fig2}(c--h).

\noindent
\textbf{Rank-1 fields.} Rank-1 fields comprise the class of coherent fields $\{\lambda_{i}\}\!=\!\{1,0,0,0\}$, and permutations thereon, whereupon $S\!=\!0$ (no statistical fluctuations). Such fields are represented by the \textit{vertices} in Fig.~\ref{fig1}(b).

\noindent
\textbf{Rank-2 fields.} Rank-2 fields where $\{\lambda_{i}\}\!=\!\{\lambda_{1},\lambda_{2},0,0\}$, and permutations thereon, correspond to the \textit{edges} of the pyramid structure [Fig.~\ref{fig1}(b)]. The entropy for rank-2 fields is $S\!=\!-\lambda_{1}\mathrm{log}_{2}\lambda_{1}-(1-\lambda_{1})\mathrm{log}_{2}(1-\lambda_{1})$,  whose value is in the range $0\!<\!S\!\leq\!1$, reaching its maximum value $S\!=\!1$ when $\lambda_{1}\!=\!\lambda_{2}\!=\!\tfrac{1}{2}$. This is a 1-parameter curve plotted in Fig.~\ref{fig2}(b), inset. Each value of entropy is associated with a unique pair of eigenvalues, so that $\mathbf{G}$ is fully identified by $S$ (modulo permutations of the eigenvalues). 

\noindent
\textbf{Rank-3 fields.} Rank-3 fields where $\{\lambda_{i}\}\!=\!\{\lambda_{1},\lambda_{2},\lambda_{3},0\}$, and permutations thereon, correspond to the \textit{faces} of the pyramid structure [Fig.~\ref{fig1}(b)] with $0\!<\!S\!\leq\!\log_{2}3$, and the maximum value of $S\!=\!\log_{2}3\!\approx\!1.585$ is reached when $\lambda_{1}\!=\!\lambda_{2}\!=\!\lambda_{3}\!=\!\tfrac{1}{3}$. In contrast to rank-2 fields, the entropy of rank-3 fields cannot uniquely identify the eigenvalues of $\mathbf{G}$ -- even after accounting for their permutations. Rather, the entropy places a constraint on the eigenvalues, thereby reducing iso-entropy rank-3 fields to a one-parameter trajectory in each face of the pyramid [Fig.~\ref{fig2}(b)]. When $S\!>\!1$, this curve is closed and contained within the triangular face and does not reach its sides. When $S\!=\!1$, the iso-entropy curve is tangential to the sides; in Fig.~\ref{fig2}(b) these tangent points are $(\lambda_{1},\lambda_{2})\!=\!(\tfrac{1}{2},\tfrac{1}{2})$, $(\tfrac{1}{2},0)$, and $(0,\tfrac{1}{2})$, which all correspond to rank-2 fields. When $S\!<\!1$, the iso-entropy locus is no longer contained within the triangular face, and instead is terminated at the sides, so that it breaks up into three unconnected curves. Ultimately this curve approaches the three vertices as $S\!\rightarrow\!0$.

\noindent
\textbf{Rank-4 fields.} For rank-4 fields, where all the eigenvalues of $\mathbf{G}$ are non-zero, we have $0\!<\!S\!\leq\!2$, and the maximum-entropy value $S\!=\!2$ is reached when all the eigenvalues are equal ($\lambda_{j}\!=\!\tfrac{1}{4}$, $j\!=\!1 .. 4$), which corresponds to a fully incoherent field. The entropy $S$ is defined over the 3D volume of the triangular pyramid [Fig.~\ref{fig1}(b)] excluding the vertices, edges and faces. Similarly to rank-3 fields, the entropy of rank-4 fields does not uniquely identify the eigenvalues of $\mathbf{G}$. The constraint placed by the entropy on the eigenvalues reduces the iso-entropy rank-4 volume to a two-parameter surface within the volume of the pyramid [Fig.~\ref{fig2}(c-f)].

Therefore, iso-entropy rank-4 fields occupy a curved \textit{surface} within this volume. When $S\!>\!1.585$, this iso-entropy surface is closed and lies entirely within the triangular pyramid. When $S\!=\!1.585$, the iso-entropy surface is tangential to all four surfaces of the triangular pyramid at their central points (which correspond to maximum-entropy rank-3 fields). When $1\!<\!S\!<\!1.585$, the iso-entropy surface intersects with each face of the triangular pyramid in a planar curve corresponding to the iso-entropy rank-3 fields having the same entropy as the rank-4 field. When $0\!<\!S\!<\!1$, the iso-entropy rank-4 surface also intersects with the edges, and these intersection points represent rank-2 fields that have the same entropy as the rank-4 field.

\subsection{Iso-entropy fields}

\begin{figure}[t!]
    \includegraphics[width=8.6cm]{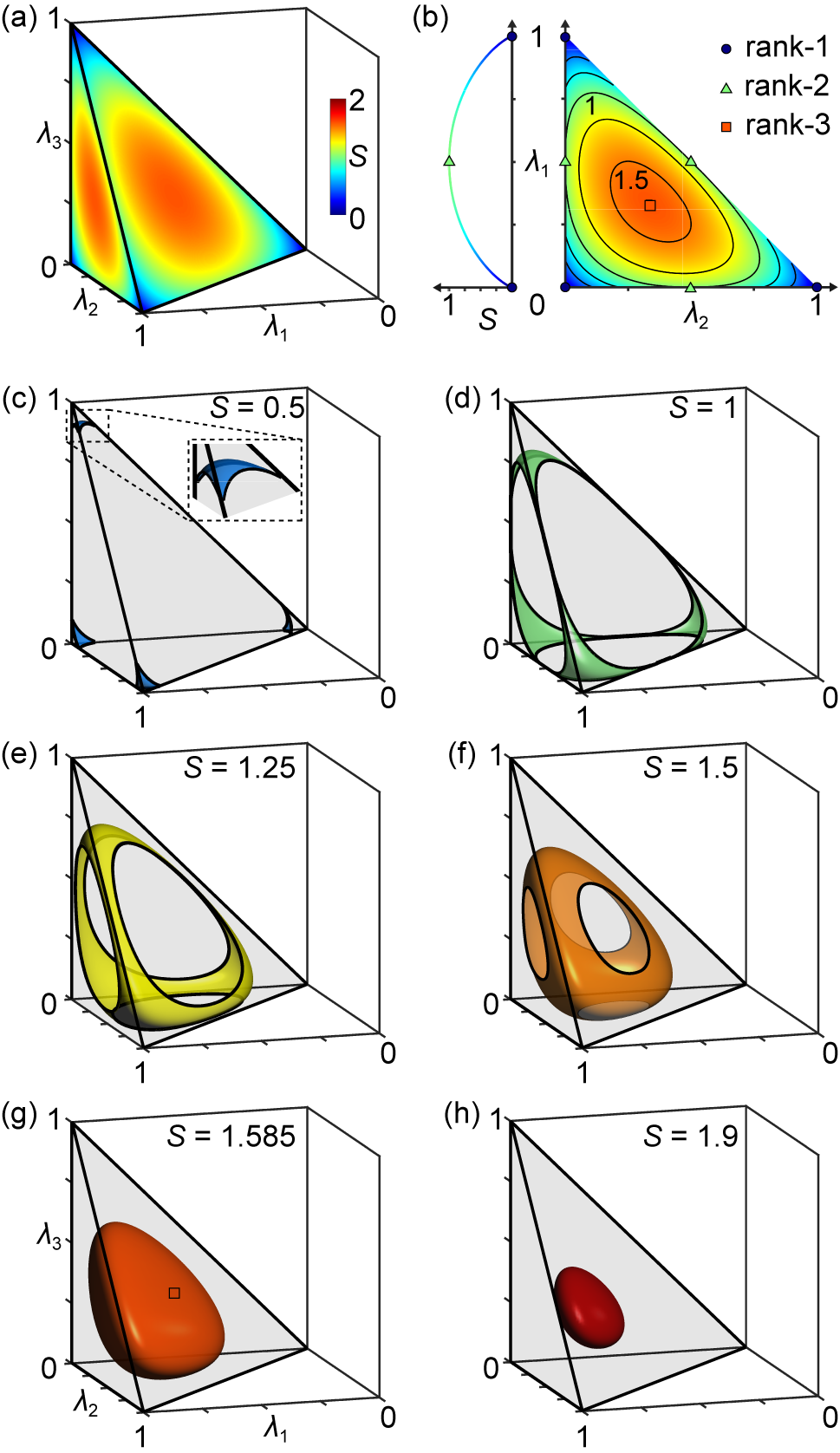}
    \caption{(a) The entropy $S$ for coherence matrices mapped onto the faces of the triangular pyramid in Fig.~\ref{fig1}(a). We show $S$ on the side face defined by $\lambda_{2}+\lambda_{3}\!=\!1$ and the front face defined by $\lambda_{1}+\lambda_{2}+\lambda_{3}\!=\!1$. (b) The entropy $S$ on the face $\lambda_{1}+\lambda_{2}\!=\!1$ from (a). Rank-1 fields correspond to the vertices at the points $(1,0)$, $(0,0)$, and $(0,1)$; rank-2 fields lie along the three sides; and rank-3 fields to the interior area. The solid contours represent iso-entropy trajectories corresponding to $S\!=\!0.5$, 0.75, 1, 1.25, and 1.5. We plot on the left $S$ for rank-2 fields in terms of $\lambda_{1}$, with $\lambda_{2}\!=\!1-\lambda_{1}$. Markers represent the maximum-entropy states for each rank; for rank-3 this is $S\!\approx\!1.585$ at the center of the triangle. (c--h) Visualization of iso-entropy surfaces corresponding to (c) $S\!=\!0.5$, (d) $S\!=\!1$, (e) $S\!=\!1.25$, (f) $S\!=\!1.5$, (g) $S\!=\!1.585$, and (h) $S\!=\!1.9$.}
    \label{fig2}
\end{figure}

In Fig.~\ref{fig2}(c-h) we examine the geometric representation of iso-entropy fields. For rank-1 fields we have $S\!=\!0$; for rank-2, $0\!<\!S\!\leq\!1$; for rank-3, $0\!<\!S\!\leq\!1.585$; and for rank-4, $0\!<\!S\!\leq\!2$. Therefore, in the range $0\!<\!S\!\leq\!1$, the field may be rank-2, rank-3, or rank-4; in the range $1\!<\!S\!\leq\!1.585$, the field may be rank-3 or rank-4; and in the range $1.585\!<\!S\!\leq\!2$, the field is exclusively rank-4.

We plot in Fig.~\ref{fig2}(c) the surface corresponding to iso-entropy fields with $S\!=\!0.5$, which comprises rank-2, rank-3, and rank-4 fields. The iso-entropy surface consists of small disconnected surfaces in the vicinity of the vertices of the triangular pyramid. Each separate area corresponds to a particular permutation of the same eigenvalues. The inset to Fig.~\ref{fig2}(c) shows an enlarged view of one of these disconnected surfaces. The portion of the \textit{surface} inside the volume represents rank-4 fields; the terminating \textit{curves} in the three neighboring faces of the pyramid represent rank-3 fields; and the terminating \textit{points} on the three edges represent rank-2 fields. 

With increase in entropy $S\!\rightarrow\!1$, the area of the iso-entropy surface increases, but the four separate areas remain disconnected. At $S\!=\!1$ the iso-entropy surface becomes a single connected surface [Fig.~\ref{fig2}(d)]. The rank-2 fields with $\mathbf{G}\!=\!\mathrm{diag}\{\tfrac{1}{2},\tfrac{1}{2},0,0\}$ lie at the points midway along the edges; the rank-3 fields correspond to the closed curves lying in each face and are tangential to the edges of the pyramid at their midpoints; and rank-4 fields correspond to the points on the surface that are within the volume.

In the range $1\!<\!S\!\leq\!1.585$, the iso-entropy surfaces no longer reach the edges (no rank-2 fields); the surface is terminated at each face with a closed curve corresponding to rank-3 fields; and the remainder of the surface inside the pyramid corresponds to rank-4 fields [Fig.~\ref{fig2}(e,f)]. As $S$ increases, the iso-entropy rank-3 curves shrink. At $S\!=\!1.585$, the iso-entropy surface is enclosed within the pyramid and is tangential to the four faces, with the tangent points corresponding to the maximum-entropy rank-3 field $\mathbf{G}\!=\!\mathrm{diag}\{\tfrac{1}{3},\tfrac{1}{3},\tfrac{1}{3},0\}$, and the remainder of the surface corresponding to rank-4 fields [Fig.~\ref{fig2}(g)]. For $1.585\!<\!S\!\leq\!2$, the iso-entropy fields in this exclusively rank-4 regime correspond to closed surfaces fully enclosed in the pyramid, not intersecting with the edges or faces [Fig.~\ref{fig2}(h)]. The size of the iso-entropy surface shrinks with increase in $S$, eventually reaching a single point $\mathbf{G}\!=\!\mathrm{diag}\{\tfrac{1}{4},\tfrac{1}{4},\tfrac{1}{4},\tfrac{1}{4}\}$ at the center of the pyramid when $S\!=\!2$.

\section{Synthesis and characterization of coherence matrices}

We proceed to describe the experimental configuration utilized for synthesizing optical fields of different rank and entropy, and for characterizing such fields via a restricted form of optical coherence matrix tomography (OCmT) \cite{Abouraddy14OL,Kagalwala15SR}.

\subsection{Synthesis of coherence matrices}

\begin{figure}[t!]
\includegraphics[width=8.2cm]{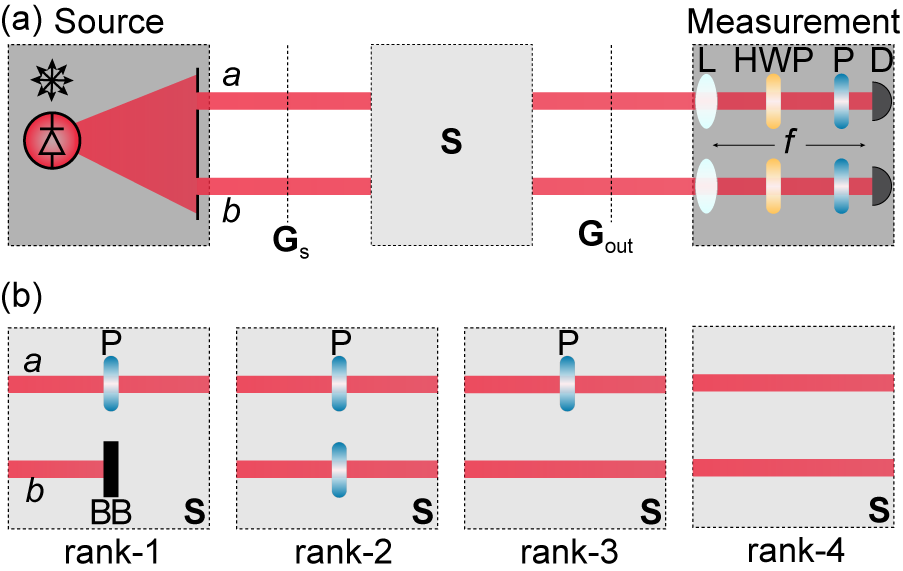}
\caption{(a) Schematic of the experimental setup for synthesizing and characterizing optical fields of different coherence rank. The transformation $\mathrm{\mathbf{S}}$ converts the source coherence matrix $\mathbf{G}_{\mathrm{s}}$ to a target $\mathbf{G}_{\mathrm{out}}$. L: Lens; HWP: half-wave plate; P: linear polarizer; and D: detector.  (b) Examples of the $\mathrm{\mathbf{S}}$ transformation for preparing fields with rank ranging from rank-1 to rank-4; BB: beam block.}
\label{fig: setup}
\end{figure}

We start with unpolarized, spatially incoherent light produced by a light-emitting diode (LED;  Thorlabs M625L4) with a center wavelength 625~nm, and a bandwidth $\approx\!17$~nm (FWHM). Two spatial modes are obtained by selecting light at two points in the field, denoted by positions `$a$' and `$b$'. This is achieved using two vertical slits of width 100~$\mu$m each, which are separated by 23~mm, which is larger than the transverse coherence width (i.e., points $a$ and $b$ are incoherent with respect to each other). We thus restrict the field to two binary DoFs: the polarization DoF spanned by the H and V polarization components, and the spatial DoF spanned by the positions $a$ and $b$ [Fig.~\ref{fig: setup}(a)]. The field produced by the source in this configuration is the maximum-entropy rank-4 field described by the coherence matrix $\mathbf{G}_{\mathrm{s}}\!=\!\mathrm{diag}\{\tfrac{1}{4},\tfrac{1}{4},\tfrac{1}{4},\tfrac{1}{4}\}$. We then make use of a \textit{non-unitary} transformation $\mathbf{S}$ to modify the field rank via projective filtering, which reduces the entropy, examples of which are depicted in Fig.~\ref{fig: setup}(b). A rank-1 field is produced by blocking the field at one point (say $b$), and placing a linear polarizer at $a$ (say along H), which results in the coherence matrix $\mathbf{G}_{\mathrm{out}}\!=\!\mathrm{diag}\{1,0,0,0\}$. A rank-2 field is produced by placing linear polarizers at both points (say along H), which yields the coherence matrix $\mathbf{G}_{\mathrm{out}}\!=\!\mathrm{diag}\{\tfrac{1}{2},0,\tfrac{1}{2},0\}$. A rank-3 field is produced by placing a linear polarizer (say along H) at one point only (say at `$b$'), which yields the coherence matrix $\mathbf{G}_{\mathrm{out}}\!=\!\mathrm{diag}\{\tfrac{1}{3},\tfrac{1}{3},\tfrac{1}{3},0\}$. Finally, a rank-4 field is produced via the identity transformation, $\mathbf{G}_{\mathrm{out}}\!=\!\mathbf{G}_{\mathrm{s}}$.

The transformations $\mathbf{S}$ depicted in Fig.~\ref{fig: setup}(b) yield the maximum-entropy field configuration for each coherence rank. Adding further optical components at $a$ and $b$ allows tuning the entropy for each rank. We have found that combinations of four optical components suffices for synthesizing an optical field of any desired rank and entropy starting from $\mathbf{G}_{\mathrm{s}}$: polarizers, half-wave plates, neutral density filters (that reduce the overall power at one spatial point with respect to the other), and partial polarizers (that adjust the power ratio of the polarization at a point). For convenience, we also made use of polarizing beam splitters, but their use is not necessary. See Supplementary Material for details. Permutations of the eigenvalues for a particular $\mathbf{G}$ can be performed by appropriate re-arrangements of the same optical components, which are provided in detail in the Supplementary Material. 

\subsection{Characterization of coherence matrices}

We reconstruct the synthesized coherence matrices using the process of OCmT \cite{Abouraddy14OL,Kagalwala15SR}, which extends to multi-DoF classical fields the well-known technique of quantum state tomography used for reconstructing multi-partite quantum states \cite{Wooters90CEPI,James01PRA1,Abouraddy02OptComm}. However, because the coherence matrices studied here are diagonalized, only 4 measurements are required, and the task of reconstructing the coherence matrix is thus simplified with respect to the general process of OCmT in which 16 measurements are required \cite{Kagalwala15SR}. If the coherence matrix is $\mathbf{G}\!=\!\mathrm{diag}\{\lambda_{1},\lambda_{2},\lambda_{3},\lambda_{4}\}$, then the eigenvalues can be determined in terms of the generalized Stokes parameters $S_{\ell m}$ that span both DoFs. To retain the nomenclature in our earlier work \cite{Abouraddy14OL,Kagalwala15SR,Harling24PRA}, the eigenvalues can be written in terms of $S_{\ell m}$ as follows:
\begin{eqnarray}
\lambda_{1}&=&S_{00}+S_{01}+S_{10}+S_{11}, \nonumber\\
\lambda_{2}&=&S_{00}-S_{01}+S_{10}-S_{11}, \nonumber\\
\lambda_{3}&=&S_{00}+S_{01}-S_{10}-S_{11}, \nonumber\\
\lambda_{4}&=&S_{00}-S_{01}-S_{10}+S_{11}.
\end{eqnarray}
These Stokes parameters in turn can be expressed in terms of measurements as follows: $S_{\ell m} = 4I_{\ell m}-2I_{0m}-2I_{\ell0}+I_{00}$, where $\ell,m = 0,1$, and the required measurements are expressed as follows $I_{00}\!=\!I_{a}+I_{b}$ (the total power); $I_{01}\!=\!I_{a\mathrm{H}}+I_{b\mathrm{H}}\!=\!I_{\mathrm{H}}$ (the total power in the H polarization component at both $a$ and $b$); $I_{10}\!=\!I_{a\mathrm{H}}+I_{a\mathrm{V}}\!=\!I_{a}$ (the total power at point $a$); and $I_{11}\!=\!I_{a\mathrm{H}}$ (the power of the H polarization component at point $a$). In all cases we made use of a power meter (Newport 843-R) connected to a silicon photodiode (Ophir, PD300R).

\begin{figure*}[t!]
    \centering
    \includegraphics[width=17.6cm]{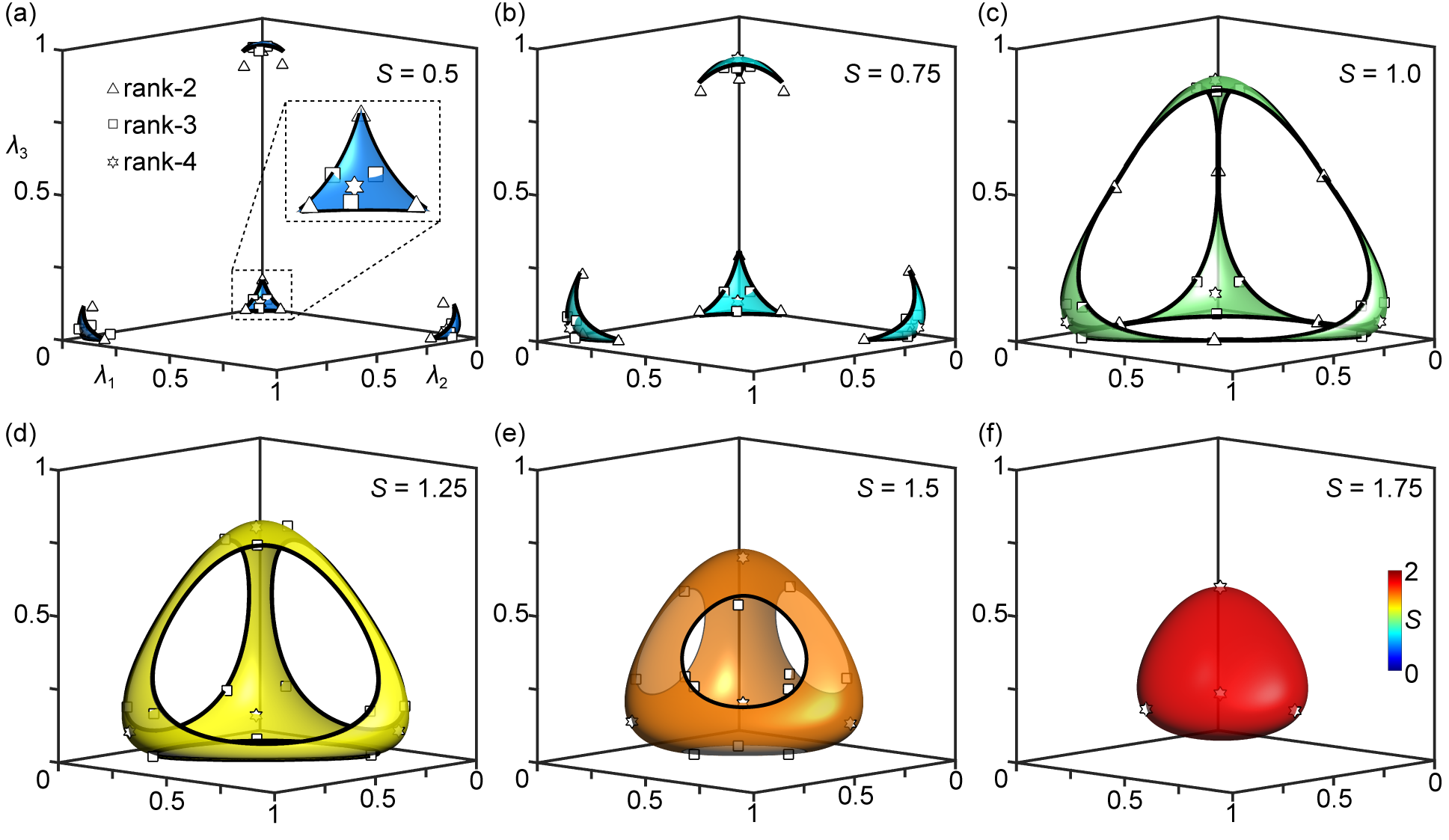}
    \caption{Experimental results for reconstructing the coherence matrices of the synthesized partially coherent optical fields of different rank represented on iso-entropy surfaces: (a) $S\!=\!0.5$, (b) $S\!=\!0.75$, (c) $S\!=\!1$, (d) $S\!=\!1.25$, (e) $S\!=\!1.5$, and (f) $S\!=\!1.75$. The inset in (a) magnifies one corner of the triangular pyramid. The white markers represent the experimentally reconstructed coherence matrices. Rank-2 fields are represented by white triangles; rank-3 fields are represented by white squares; and rank-4 fields are represented by white hexagrams.}
    \label{fig:GeometricDomains}
\end{figure*}

\section{Measurement results}

We plot in Fig.~\ref{fig:GeometricDomains} the OCmT measurement results for a wide range of $4\times4$ matrices of different rank and entropy values. For each value of entropy, we synthesize and characterize optical fields whose rank is compatible with $S$. We have excluded zero-entropy rank-1 fields.

When $S\!=\!0.5$ [Fig.~\ref{fig:GeometricDomains}(a)], we produce fields with coherence rank 2, 3, and 4 corresponding to the coherence matrices $\mathbf{G}^{(2)}$, $\mathbf{G}^{(3)}$, and $\mathbf{G}^{(4)}$, respectively:
\begin{eqnarray}
\mathbf{G}^{(2)}&=&\mathrm{diag}\{0.89, 0.11, 0, 0\},\nonumber\\
\mathbf{G}^{(3)}&=&\mathrm{diag}\{0.916, 0.042, 0.042,0\},\nonumber\\
\mathbf{G}^{(4)}&=&\mathrm{diag}\{0.925,0.025, 0.025, 0.025\}.
\end{eqnarray}
For $\mathbf{G}^{(2)}$, we have experimentally synthesized and reconstructed 12 distinct permutations, for $\mathbf{G}^{(3)}$ there are 12, and for $\mathbf{G}^{(4)}$ there are 4. The points corresponding to the 28 reconstructions via OCmT are plotted in Fig.~\ref{fig:GeometricDomains}(a). As noted earlier, the iso-entropy surface $S\!=\!0.5$ comprises 4 disconnected areas, with one of these areas expanded in the inset of Fig.~\ref{fig:GeometricDomains}(a). Here, 3 permutations for the rank-2 field are at the vertices of the area (the intersection points with the edges of the pyramid); 3 permutations of the rank-3 field lie on the curved edges of the area (which lie on the faces of the pyramid); and the rank-4 field lies on the area (which lies within the pyramid volume). The remaining permutations of the coherence matrices lie on the other 3 disconnected areas in Fig.~\ref{fig:GeometricDomains}(a).

When $S\!=\!0.75$ [Fig.~\ref{fig:GeometricDomains}(b)], fields with coherence rank 2, 3, and 4 can be produced with targeted coherence matrices $\mathbf{G}^{(2)}$, $\mathbf{G}^{(3)}$, and $\mathbf{G}^{(4)}$, respectively:
\begin{eqnarray}
\mathbf{G}^{(2)}&=&\mathrm{diag}\{0.785, 0.215, 0, 0\},\nonumber\\
\mathbf{G}^{(3)}&=&\mathrm{diag}\{0.852, 0.074, 0.074,0\},\nonumber\\
\mathbf{G}^{(4)}&=&\mathrm{diag}\{0.873,0.042, 0.042, 0.042\}.
\end{eqnarray}
The experimentally realized coherence matrices have the same 28 permutations as those above for $S\!=\!0.5$, with a similar distribution across the 4 disconnected areas in Fig.~\ref{fig:GeometricDomains}(b). 

When $S\!=\!1.0$ 
[Fig.~\ref{fig:GeometricDomains}(c)], fields with coherence rank 2, 3, and 4 can be produced with targeted coherence matrices $\mathbf{G}^{(2)}$, $\mathbf{G}^{(3)}$, and $\mathbf{G}^{(4)}$, respectively:
\begin{eqnarray}
\mathbf{G}^{(2)}&=&\mathrm{diag}\{0.5,0.5, 0, 0\},\nonumber\\
\mathbf{G}^{(3)}&=&\mathrm{diag}\{0.772, 0.114, 0.114, 0\}.,\nonumber\\
\mathbf{G}^{(4)}&=&\mathrm{diag}\{0.811,0.063, 0.063, 0.063\}.
\end{eqnarray}
For $\mathbf{G}^{(2)}$ there are 6 distinct permutations, for $\mathbf{G}^{(3)}$ there are 12, and for $\mathbf{G}^{(4)}$ there are 4. The points, corresponding to the 22 reconstructions via OCmT, are plotted in Fig.~\ref{fig:GeometricDomains}(c). At $S\!=\!1$ the iso-entropy surface is for the first time connected. The permutations of $\mathbf{G}^{(2)}$ are the 6 mid-points along the edges of the pyramid. This iso-entropy surface is tangential to the pyramid at these points. The permutations of $\mathbf{G}^{(3)}$ lie along the curved edges of this surface, which all lie on the faces of the pyramid. The permutations of $\mathbf{G}^{(4)}$ lie on the iso-entropy surface away from its edges, which lies within the pyramid.

When $S\!=\!1.25$ [Fig.~\ref{fig:GeometricDomains}(d)], fields with coherence rank 3 and 4 can be produced with targeted coherence matrices $\mathbf{G}^{(3)}$ and $\mathbf{G}^{(4)}$, respectively:
\begin{eqnarray}
\mathbf{G}^{(3)}&=&\mathrm{diag}\{0.668, 0.166, 0.166, 0\},\nonumber\\
\mathbf{G}^{(4)}&=&\mathrm{diag}\{0.736, 0.088, 0.088, 0.088\}.
\end{eqnarray}
For $\mathbf{G}^{(3)}$ there are 12 distinct permutations, and for $\mathbf{G}^{(4)}$ there are 4. The points corresponding to the 16 reconstructions via OCmT are plotted in Fig.~\ref{fig:GeometricDomains}(d). 

For $S\!=\!1.5$ [Fig.~\ref{fig:GeometricDomains}(e)], fields with coherence rank 3 and 4 can be produced with targeted coherence matrices $\mathbf{G}^{(3)}$ and $\mathbf{G}^{(4)}$, respectively:
\begin{eqnarray}
\mathbf{G}^{(3)}&=&\mathrm{diag}\{0.5, 0.25, 0.25, 0\},\nonumber\\
\mathbf{G}^{(4)}&=&\mathrm{diag}\{0.646, 0.118, 0.118, 0.118\}.
\end{eqnarray}
There are 12 distinct permutations for $\mathbf{G}^{(3)}$, and for $\mathbf{G}^{(4)}$ there are 4. The points corresponding to the 16 reconstructions via OCmT are plotted in Fig.~\ref{fig:GeometricDomains}(e), and are similar to those in Fig.~\ref{fig:GeometricDomains}(d) except for the more compact area of the iso-entropy surface. 

Finally, when $S\!=\!1.75$ [Fig.~\ref{fig:GeometricDomains}(f)], only fields with coherence rank 4 can be produced with target coherence matrix $\mathbf{G}^{(4)}$:
\begin{equation}
\mathbf{G}^{(4)}=\mathrm{diag}\{0.526, 0.158, 0.158, 0.158\}.
\end{equation}
There are 4 distinct permutations for $\mathbf{G}^{(4)}$, and the points corresponding to the 4 reconstructions via OCmT are plotted in Fig.~\ref{fig:GeometricDomains}(f).

In the Supplementary Material, we provide the optical arrangement for synthesizing each coherence matrix, the measured eigenvalues, and a comparison to the representation of the theoretical coherence matrices on the iso-entropy surfaces corresponding to Fig.~\ref{fig:GeometricDomains}.

\section{Steering the coherence matrix across an intra-rank iso-entropy trajectory}

\begin{figure}[t!]
    \includegraphics[width=8.2cm]{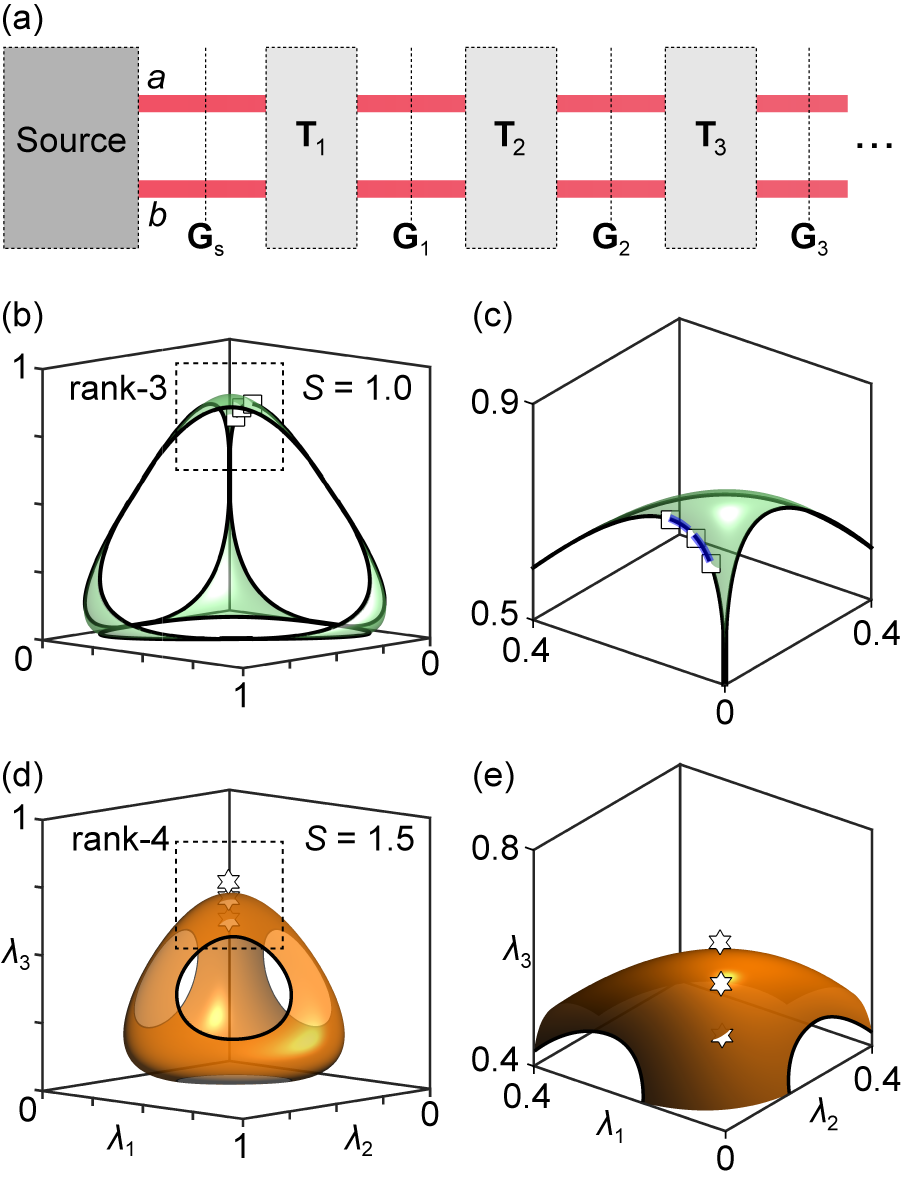}
    \caption{(a) The general configuration for steering the coherence matrix along iso-entropy intra-rank and inter-rank trajectories in the geometric space depicted in Fig.~\ref{fig1}(a). The transformations $\mathrm{\mathbf{T}}_j$ are non-unitary. The coherence matrices $\mathbf{G}_{j}$ are reconstructed after each transformation $\mathbf{T}_{j}$. (b) Experimental results for steering rank-3 coherence matrices along the an intra-rank iso-entropy \textit{curve} corresponding to $S\!=\!1$. (c) A zoomed in and rotated view of the portion of (b) enclosed in the dotted black square. (d,e) Same as (b) but for rank-4 fields along an intra-rank iso-entropy \textit{surface} corresponding to $S\!=\!1.5$.}
    \label{fig:intra-rank}
\end{figure}

Each of the 114 coherence matrices described above was synthesized directly from the maximum-entropy, rank-4 source coherence matrix $\mathbf{G}_{\mathrm{s}}\!=\!\mathrm{diag}\{\tfrac{1}{4},\tfrac{1}{4},\tfrac{1}{4},\tfrac{1}{4}\}$ via a distinct optical arrangement that determines the rank and entropy of the synthesized field. Here we consider a different scenario where we start from the source $\mathbf{G}_{\mathrm{s}}$, but then subsequently apply a sequence of transformation $\mathbf{T}_{j}$ [Fig.~\ref{fig:intra-rank}(a)] to steer the coherence matrix over an iso-entropy trajectory on one of the iso-entropy surfaces [Fig.~\ref{fig2}]. After each such transformation, the coherence matrix is reconstructed. We consider here this strategy for field transformations that maintain the coherence rank, which we denote `intra-rank' iso-entropy transformations.

Given a particular coherence matrix $\mathbf{G}$, one could of course apply a myriad of unitaries that produce different field configurations, but their coherence matrices are represented by the same point in the space defined in Fig.~\ref{fig1}(a). We thus exclude these unitaries here. A special case of unitaries are those that produce permutations of the eigenvalues of $\mathbf{G}$ in the diagonal representation. Although the corresponding points representing these different diagonal representations are distinct (as shown in Fig.~\ref{fig:GeometricDomains}), we also exclude these transformations here since the eigenvalues remain invariant. This exhausts all the possibilities for rank-1 and rank-2 fields where it is always possible to perform intra-rank conversion between iso-entropy fields unitarily. We are thus concerned here with transformations that maintain the rank \textit{and} the entropy for rank-3 and rank-4 fields, but change the eigenvalues. These are necessarily \textit{non}-unitary transformations.

We first steer the coherence matrix of rank-3 fields along an iso-entropy curve at $S\!=\!1$. Starting from the source coherence matrix $\mathbf{G}_{\mathrm{s}}$, we first transform $\mathbf{G}_{\mathrm{s}}$ to $\mathbf{G}_{1}$, $\mathbf{G}_{1}$ to $\mathbf{G}_{2}$, and then $\mathbf{G}_{2}$ to $\mathbf{G}_{3}$:
\begin{equation}
\mathbf{G}_{\mathrm{s}}\xrightarrow{\mathbf{T}_{1}}\mathbf{G}_{1}\xrightarrow{\mathbf{T}_{2}}\mathbf{G}_{2}\xrightarrow{\mathbf{T}_{3}}\mathbf{G}_{3},
\end{equation}
where the reconstructed coherence matrices are:
\begin{eqnarray}
\mathbf{G}_{1}&=&\mathrm{diag}\{0.03, 0, 0.72, 0.25\},\nonumber\\
\mathbf{G}_{2}&=&\mathrm{diag}\{0.06, 0, 0.75, 0.19\},\nonumber\\
\mathbf{G}_{3}&=&\mathrm{diag}\{0.11, 0, 0.77, 0.11\},
\end{eqnarray}
and the requisite entropy-preserving non-unitary transformations $\mathbf{T}_{1}$, $\mathbf{T}_{2}$, and $\mathbf{T}_{3}$ are:
\begin{eqnarray}
\mathbf{T}_{1}&=&\mathrm{diag}\{0.205, 0, 1, 0.597\},\nonumber\\
\mathbf{T}_{2}&=&\mathrm{diag}\{1, 1, 0.726, 0.602\},\nonumber\\
\mathbf{T}_{3}&=&\mathrm{diag}\{1, 1, 0.734, 0.568\}.
\end{eqnarray}
These transformations can be constructed from partial polarizers and neutral density filters placed at $a$ and $b$. A partial polarizer at $a$ would reduce the overall entropy, but then reducing the power at $a$ with respect to $b$ can counter-balance this decrease, and thus return the entropy to its initial value. The optical configurations corresponding to each of these non-unitary transformations are provided in the Supplementary Material. The points corresponding to the experimentally reconstructed coherence matrices represented in $\{\lambda_{1},\lambda_{2},\lambda_{3}\}$-space are plotted in Fig.~\ref{fig:intra-rank}(b,c); the corresponding theoretically targeted coherence matrices are plotted in the Supplementary Material.

Next, we steer the coherence matrix along an iso-entropy trajectory on the iso-entropy rank-4 surface at $S\!=\!1.5$ while maintaining the coherence rank at 4. Starting from the source coherence matrix $\mathbf{G}_{\mathrm{s}}$, we convert it to the coherence matrix $\mathbf{G}_{4}$, $\mathbf{G}_{4}$ to $\mathbf{G}_{5}$, and then $\mathbf{G}_{5}$ to $\mathbf{G}_{6}$:
\begin{equation}
\mathbf{G}_{\mathrm{s}}\xrightarrow{\mathbf{T}_{4}}\mathbf{G}_{4}\xrightarrow{\mathbf{T}_{5}}\mathbf{G}_{5}\xrightarrow{\mathbf{T}_{6}}\mathbf{G}_{6},
\end{equation}
and the reconstructed coherence matrices are:
\begin{eqnarray}
\mathbf{G}_{4}&=&\mathrm{diag}\{0.12, 0.13, 0.64, 0.11\},\nonumber\\
\mathbf{G}_{5}&=&\mathrm{diag}\{0.07, 0.09, 0.61, 0.22\},\nonumber\\
\mathbf{G}_{6}&=&\mathrm{diag}\{0.05, 0.08, 0.53, 0.35\},
\end{eqnarray}
which are produced by the non-unitary transformations $\mathbf{T}_{4}$, $\mathbf{T}_{5}$, and $\mathbf{T}_{6}$ given by:
\begin{eqnarray}
\mathbf{T}_{4}&=&\mathrm{diag}\{0.428, 0.428, 1, 0.428\},\nonumber\\
\mathbf{T}_{5}&=&\mathrm{diag}\{0.595, 0.595, 0.704, 1\},\nonumber\\
\mathbf{T}_{6}&=&\mathrm{diag}\{0.701, 0.701, 0.755, 1\}.
\end{eqnarray}
The optical configurations corresponding to each of these non-unitary transformations are provided in the Supplementary Material. The points corresponding to the experimentally reconstructed coherence matrices represented in $\{\lambda_{1},\lambda_{2},\lambda_{3}\}$-space are plotted in Fig.~\ref{fig:intra-rank}d,ec); the corresponding theoretically targeted coherence matrices are plotted in the Supplementary Material.

\section{Steering the coherence matrix across an inter-rank iso-entropy trajectory}

\begin{figure}[t!]
    \includegraphics[width=8.2cm]{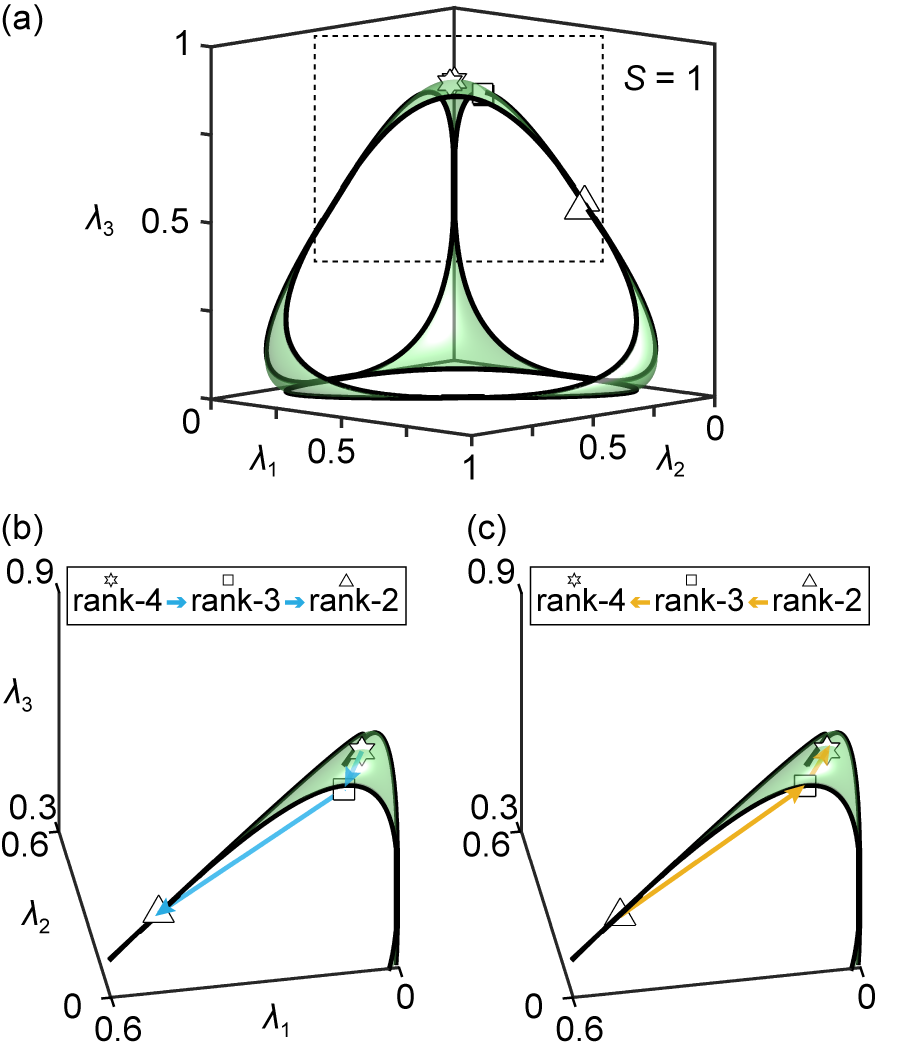}
    \caption{(a) Experimental results for \textit{inter}-rank steering of a coherence matrix along an iso-entropy trajectory corresponding to $S\!=\!1$. (b) Zoomed-in section of the plot in (a) showing the results of the descending order trajectory from rank-4 to rank-3 to rank-2. (c) Same as (b) but for ascending order from rank-2 to rank-3 to rank-4.}
    \label{fig:inter-rank}
\end{figure}

We now consider iso-entropy trajectories that extend across ranks, which we denote `inter-rank' transformations. These also require entropy-preserving non-unitary transformations for their realization. We carry out our experiments over an iso-entropy inter-rank trajectory with $S\!=\!1$. Starting with the rank-4 source, we first convert $\mathbf{G}_{\mathrm{s}}$ to $\mathbf{G}_{4}$ (rank-4), $\mathbf{G}_{4}$ to $\mathbf{G}_{3}$ (rank-3), and then $\mathbf{G}_{3}$ to $\mathbf{G}_{2}$ (rank-2), where these iso-entropy coherence matrices are given by:
\begin{equation}
\mathbf{G}_{\mathrm{s}}\xrightarrow{\mathbf{T}_{1}}\mathbf{G}_{4}\xrightarrow{\mathbf{T}_{2}}\mathbf{G}_{3}\xrightarrow{\mathbf{T}_{3}}\mathbf{G}_{2},
\end{equation}
The required entropy-preserving non-unitary transformations utilized to produce these changes are given by:
\begin{eqnarray}
\mathbf{T}_{1}&=&\mathrm{diag}\{0.279, 0.279, 1, 0.279\},\nonumber\\
\mathbf{T}_{2}&=&\mathrm{diag}\{1, 0, 0.727, 1\},\nonumber\\
\mathbf{T}_{3}&=&\mathrm{diag}\{1, 1, 0.383, 0\}.
\end{eqnarray}
These transformations can be constructed from partial polarizers and neutral density filters placed at $a$ and $b$. The reconstructed coherence matrices are given by:
\begin{eqnarray}
\mathbf{G}_{4}&=&\mathrm{diag}\{0.06, 0.07, 0.81, 0.06\},\nonumber\\
\mathbf{G}_{3}&=&\mathrm{diag}\{0.11, 0, 0.76, 0.12\},\nonumber\\
\mathbf{G}_{2}&=&\mathrm{diag}\{0.5, 0, 0.5, 0\}.
\end{eqnarray}
The points corresponding to the experimentally reconstructed coherence matrices represented in $\{\lambda_{1},\lambda_{2},\lambda_{3}\}$-space are plotted in Fig.~\ref{fig:inter-rank}(a,b).

We now consider traversing the \textit{same} iso-entropy ($S\!=\!1$) trajectory in the opposite direction:
\begin{equation}
\mathbf{G}_{4}\xleftarrow{\mathbf{T}_{6}}\mathbf{G}_{3}\xleftarrow{\mathbf{T}_{5}}\mathbf{G}_{2}\xleftarrow{\mathbf{T}_{4}}\mathbf{G}_{\mathrm{s}}.
\end{equation}
The transformation $\mathbf{T}_{4}$ that converts the source $\mathbf{G}_{\mathrm{s}}$ to the rank-2 $\mathbf{G}_{2}$ is given by:
\begin{equation}
\mathbf{T}_{4}=\mathrm{diag}\{1, 0, 1, 0\},
\end{equation}
which involves filtering out the V polarization component at $a$ and $b$ (resulting in a field that is spatially incoherent and linearly polarized along H). Converting $\mathbf{G}_{2}$ to $\mathbf{G}_{3}$ poses a new challenge: whereas \textit{reducing} the coherence rank can always be done by projective filtering, \textit{increasing} the coherence rank cannot. Instead, to introduce a third non-zero eigenvalue as required here necessitates a \textit{randomizing} transformation rather than a deterministic one, which invariably increases the field entropy. Following such a transformation with an appropriate projective filter restores the entropy to the target value without changing the coherence rank.

We implement this randomizing non-unitary transformation by placing a rotating HWP at $a$ and averaging the detection over multiple full rotations, which has the effect of randomizing the polarization at $a$. The trasnsformation corresponding to this randomizing system cannot be described by a $4\!\times\!4$ operator $\mathbf{T}$, and instead requires a `super-operator' representation (see Supplementary Material). We make use of two such polarization randomizers, one at $a$ and the other at $b$ to increase the rank from 2 to 3, and then from 3 to 4, respectively. Each HWP rotates at $25^{\circ}$/s, and measurements are averaged over 30~s. Following this randomizing operation is a projective filtering operation to restore the target entropy. The transformation $\mathbf{T}_{5}$ comprises a polarization randomizer placed at $a$ followed by a projective filter $\mathbf{T}_{5}'$, and $\mathbf{T}_{6}$ comprises a polarization randomizer placed at $b$ followed by a projective filter $\mathbf{T}_{6}'$, where $\mathbf{T}_{5}'$ and $\mathbf{T}_{6}'$ are given by:
\begin{eqnarray}
\mathbf{T}_{5}'&=&\mathrm{diag}\{0.271, 0.271, 1, 0.383\},\nonumber\\
\mathbf{T}_{6}'&=&\mathrm{diag}\{1, 1, 0.972, 0.707\}.
\end{eqnarray}
The reconstructed coherence matrices are given by:
\begin{eqnarray}
\mathbf{G}_{2}&=&\mathrm{diag}\{0.50, 0.02, 0.48, 0\},\nonumber\\
\mathbf{G}_{3}&=&\mathrm{diag}\{0.11, 0.01, 0.77, 0.11\},\nonumber\\
\mathbf{G}_{4}&=&\mathrm{diag}\{0.06, 0.08, 0.80, 0.07\}.
\end{eqnarray}
The points corresponding to the experimentally reconstructed coherence matrices represented in $\{\lambda_{1},\lambda_{2},\lambda_{3}\}$-space are plotted in Fig.~\ref{fig:inter-rank}(a,c). See Supplementary Material for more detail on the non-unitary transformations used for iso-entropy inter-rank conversions.

\section{Discussion}

\subsection{Relevance to optical communications}

The formulation of optical coherence presented here (in terms of multiple coupled discrete DoFs) immediately suggests applications in optical communications and sensing, whereupon one encodes information physically into the eigenvalues of the joint coherence matrix $\mathbf{G}$. Specifically, such a scheme offers unique advantages with regards to propagation in a perturbing environment. Multiple models can be adopted for the environment:

\textit{1. Unitary transformation of the polarization and/or the spatial DoFs:} Such a transformation of course changes the polarization state and/or the spatial state. However, these transformations do \textit{not} change the degree of coherence of either DoF (after tracing out the other DoF). Additionally, a medium modeled by such a transformation leaves the eigenvalues of $\mathbf{G}$ invariant. 

\textit{2. Unitary coupling of the two DoFs:} Such a transformation of course changes the polarization and spatial states, and also changes the degree of coherence of both DoFs \cite{Abouraddy17OE}. Nevertheless, such a transformation does \textit{not} affect the eigenvalues of $\mathbf{G}$.

\textit{3. Non-unitary transformation of the DoFs:} The coherence matrix is immune to certain classes of non-unitary transformations, such as an overall lossy channel that reduces the power of all the field modes equally. Such a transformation does not change the eigenvalues of $\mathbf{G}$. However, $\mathbf{G}$ will be distorted through other non-unitary transformations, such as (1) polarization-dependent losses or spatially dependent losses; (2) projective filtering transformations (that reduce the entropy); and (3) randomizing channels (that increase the entropy). 

We have \textit{not} considered here optical channels that introduce additive noise sources that are independent of $\mathbf{G}$, which will of course change its eigenvalues. It would be interesting to consider the impact of such sources of noise, in addition to the non-unitary transformations mentioned above, on the representation of a coherence matrix in the geometric space depicted in Fig.~\ref{fig1}. This would inform the choice of coherence matrices whose separating `distance' in this space is sufficient to render their states of the associated optical fields sufficiently distinct after transmission through such a prescribed channel. Much experimental and theoretical work is anticipated along these lines.


\subsection{Larger-dimensional DoFs}

We have couched our formulation in terms of two binary DoFs. However, this analysis can be readily extended to larger-dimensional DoFs. In general, for two DoFs of dimension $N_{1}$ and $N_{2}$, the dimension of the joint space is $N_{1}\times N_{2}$, and the size of the associated coherence matrix is $(N_{1}\times N_{1})\times(N_{1}\times N_{2})$. Although this provides a large-dimensional space to increase the information-carrying capacity, the number of measurements required to reconstruct $\mathbf{G}$ nevertheless poses a challenge, which we address below.

An example is transmission over a multimode fiber. One may easily increase the number of available modes by increasing the fiber diameter. In so-called spatial-mode multiplexing, the number of channels in a multimode fiber is increased over that in a single-mode fiber by exploiting each spatial mode (in an orthogonal modal set) as an independent communications channel \cite{Richardson13NP,Li14AOP,Willner15AOP,Willner21APR,Wang23JLT}. A major impediment for such schemes is of course the potential coupling between these modes at fiber bends, or caused by variations in temperature or stress in the fiber, especially over large propagation distances. Moreover, the spatial modes may couple to polarization \cite{Vitullo17PRL}, which can further distort the communications channels. Our approach here could help address this challenge by encoding the information globally in $\mathbf{G}$.

In addition to the spatial modes utilized here or those of a multi-mode optical fiber, our scheme can be readily extended to other spatial modal bases, including those of orbital angular momentum \cite{Kagalwala15SR, zou2022tunability}, spatial parity states \cite{Abouraddy07PRA,Yarnall07PRL,Kagalwala13NP}, among a host of others \cite{Diouf21OE,yessenov2022space,Yessenov22AOP,Yessenov22OL}. In principle, the same approach outlined here can be extended to quantum states of light, whether single-photon or entangled-photon states \cite{Kagalwala17NC}.

\subsection{The potential for high-speed optical communications with partially coherent fields}

Of course, a high-speed automated approach to reconstructing the coherence matrix is needed to make such schemes relevant to optical communications. The same constraint applies to the synthesis of partially coherent fields with a given coherence matrix. In this context, a recent proposal by Miller~\textit{et al.} \cite{roques2024measuring} opens a new avenue that increases the relevance of partially coherent light for optical communications. In earlier work on \textit{coherent} fields, the independent communication modes were determined that can be established between arbitrary transmitters and receivers defined solely by the relative positions of their two volumes \cite{Miller98OL,Miller2000AO,Miller19AOP}. Subsequently, a generic algorithm was proposed that could be used to identify these channels without prior knowledge, and an implementation in terms of a mesh of Mach-Zehnder interferometers was put forth \cite{Miller13JLT,Miller13PR}. This allows for realizations of the entire procedure to be integrated onto a photonic chip \cite{Bogaerts20Nature}. The recent theoretical proposal in \cite{roques2024measuring} extends this strategy to partially coherent fields, and the independent communication channels can be identified with the basis of the diagonalized coherence matrix. Our experiments here deal with two binary DoFs rather than one high-dimensional DoF, but the general strategy in \cite{roques2024measuring} can likely be modified to adapt to the multi-DoF configuration. Importantly, the on-chip realization proposed in \cite{roques2024measuring} would make possible high-speed synthesis and measurements of these multi-DoF partially coherent fields, which may stand to revolutionize the applications of such fields.   

Finally, we note a distinct theoretical proposal made recently by Novotny~\textit{et al}. \cite{Nardi22OL}, in which a partially coherent field with a high-dimensional spatial DoF is used for enhancing the channel capacity over a multi-mode optical fiber by encoding information in the correlations between the various basis modes (corresponding to the off-diagonal elements of $\mathbf{G}$). This is a distinct proposal for utilizing the partial coherence of an optical field to carry out a task for which a coherent field of the same dimension falls short. Such a scheme is susceptible to scattering, in contrast to our proposal here that is immune to scattering (as defined above). However, our approach does \textit{not} offer a higher channel capacity as in \cite{Nardi22OL}. These recent developments indicate the growing awareness of the rich possibilities made possible with partially coherent light that are only now coming to the fore.

\section{Conclusions}

In conclusion, we have explored the geometry of iso-entropy, partially coherent optical fields comprising two binary DoFs -- polarization and a pair of spatial modes. In the case of optical fields characterized by a single binary DoF, any two iso-entropy field configurations can always be converted into each other via a unitary transformation. In contrast, iso-entropy optical fields combining two such DoFs -- rather than one -- do not follow the same pattern. Instead, the rank of the associated $4\times4$ coherence matrix -- the number of its non-zero eigenvalues -- plays a deciding role. The entropy for rank-1 and rank-2 fields uniquely determines the optical fields that can be converted into each other unitarily. This is \textit{not} the case for rank-3 or rank-4 fields. To convert two iso-entropy rank-3 fields (or two iso-entropy rank-4 fields) into each other, one may need to resort to non-unitary transformations. Moreover, inter-rank transformations can only be achieved using non-unitary systems.

We have experimentally synthesized a wide range of partially coherent fields of different rank and entropy, and have tomographically reconstructed their associated coherence matrices. Finally, we have steered the coherence matrix over intra-rank and inter-rank iso-entropy trajectories via non-unitary transformations. These results suggest new applications for partially coherent light in optical communications schemes that may offer advantages in the presence of scattering in the optical channel.

\noindent
\textit{Acknowledgments}. We thank Aristide Dogariu for useful discussions and Joshua A. Burrow for assistance preparing the figures. This work was funded by the US Office of Naval Research (ONR) under contracts N00014-17-1-2458 and N00014-20-1-2789.

\bibliography{sample}

\end{document}


\preprint{APS/123-QED}

\title{Supplementary: Iso-entropy partially coherent optical fields that cannot be inter-converted unitarily}

\author{Mitchell Harling}
\affiliation{PROBE Lab, School of Engineering, Brown University, Providence, RI 02912, USA}%
\author{Varun A. Kelkar}
\affiliation{Department of Electrical and Computer Engineering, University of Illinois at Urbana-Champaign, Urbana, IL 61801, USA}%
\affiliation{Currently at Algorithmic Systems Group, Analog Garage, Analog Devices, Inc., Boston, MA 02110, USA}
\author{Kimani C. Toussaint, Jr.}
\affiliation{PROBE Lab, School of Engineering, Brown University, Providence, RI 02912, USA}%
\affiliation{Brown-Lifespan Center for Digital Health, Providence, RI 02912, USA}%
\author{Ayman F. Abouraddy}%
\affiliation{CREOL, The College of Optics \& Photonics, University of Central Florida, Orlando, FL 32816, USA}%

\date{\today}

\begin{abstract}

\end{abstract}

\maketitle

\section{Operator representation of the optical devices used}

In the diagonal representation of the coherence matrix $\mathbf{G}$, we need only use `local' operators to modify its structure. By local operators we mean optical devices that are placed in the path of $a$ and $b$, with no need to mix the fields from $a$ and $b$. Therefore, the transformation of $\mathbf{G}$ in this scenario takes the form $\mathbf{G}'\!=\!\hat{V}\mathbf{G}\hat{V}^{\dag}$, where the transformation $\hat{V}$ (not necessarily unitary) has a block diagonal form: $\hat{V}\!=\!\left(\begin{array}{cc}\hat{V}_{a}&\mathbf{0}\\\mathbf{0}&\hat{V}_{b}\end{array}\right)$, where $\mathbf{0}$ is a $2\times2$ matrix of zeros, and $\hat{V}_{a}$ and $\hat{V}_{b}$ are $2\times2$ matrices representing the polarization-sensitive or insensitive devices placed at $a$ and $b$. The devices used in our work here include the following: 
\begin{enumerate}

\item \textit{Linear polarizer} P: The operator for projecting polarization on a linear basis rotated by an angle $\theta$ with respect to H is given by $\hat{V}_{\mathrm{P}}(\theta)\!=\!\left(\begin{array}{cc}\cos^{2}\theta&\cos\theta\sin\theta\\\cos\theta\sin\theta&\sin^{2}\theta\end{array}\right)$. For a linear polarizer along H we have $\hat{V}_{\mathrm{H}}\!=\!\hat{V}_{\mathrm{P}}(0)\!=\!\left(\begin{array}{cc}1&0\\0&0\end{array}\right)$; and for a linear polarizer along V we have $\hat{V}_{\mathrm{V}}\!=\!\hat{V}_{\mathrm{P}}(90^{\circ})\!=\!\left(\begin{array}{cc}0&0\\0&1\end{array}\right)$.

\item \textit{Partial polarizer} (PP): $\hat{V}_{\mathrm{PP}}(\alpha_{\mathrm{H}}, \alpha_{\mathrm{V}})\!=\!\left(\!\begin{array}{cc}\alpha_{\mathrm{H}}&0\\0&\alpha_{\mathrm{V}}\end{array}\!\right)$, where $\alpha_{\mathrm{H}}$ and $\alpha_{\mathrm{V}}$ are the attenuation factors for the H and V polarization components, respectively.

\item \textit{Neutral density filter} (ND): $\hat{V}_{\mathrm{ND}}(\alpha)\!=\!\left(\!\begin{array}{cc}\alpha&0\\0&\alpha\end{array}\!\right)$, where $\alpha$ is the polarization-independent attenuation factor corresponding to the optical density of the ND filter.

\item \textit{Half-wave plate} (HWP): $\hat{V}_{\mathrm{HWP}}(\phi)\!=\!\left(\begin{array}{cc}
\cos2\phi&\sin2\phi\\\sin2\phi&-\cos2\phi\end{array}\!\right)$, where $\phi$ is the angle of the fast axis of the HWP with respect to H.

\item \textit{Beam block} (BB): $\hat{V}_{\mathrm{BB}}\!=\!\left(\begin{array}{cc}
0&0\\0&0\end{array}\!\right)\!=\!\mathbf{0}$ that blocks $100 \%$ of the light.

\item \textit{Polarizing beam splitter} (PBS): This is the sole device used that mixes light from $a$ and $b$. For a PBS that transmits horizontally polarized light and reflects vertically polarized light, the matrix representation is:
\begin{equation}
\hat{V}_{\mathrm{PBS}}=\left(\begin{array}{cccc}
0&0&1&0\\
0&i&0&0\\
1&0&0&0\\
0&0&0&i\end{array}\!\right).
\end{equation}
\end{enumerate}
The operators $\hat{V}_{\mathrm{P}}(\theta)$, $\hat{V}_{\mathrm{PP}}(\alpha_{\mathrm{H}},\alpha_{\mathrm{V}})$, $\hat{V}_{\mathrm{ND}}(\alpha)$, and $\hat{V}_{\mathrm{BB}}$ correspond to deterministic \textit{non-unitary} transformations, whereas $\hat{V}_{\mathrm{HWP}}(\phi)$ and $\hat{V}_{\mathrm{PBS}}$ correspond to deterministic \textit{unitary} transformations.

\noindent\textbf{Randomizing the polarization DoF via a rotating HWP}

The goal is to establish an optical system that randomizes the state of polarization at $a$ or $b$ regardless of its initial state. Consider a polarization coherence matrix at a point in the optical field:
\begin{equation}
\mathbf{G}=\left(\begin{array}{cc}G_{\mathrm{HH}}&G_{\mathrm{HV}}\\G_{\mathrm{VH}}&G_{\mathrm{VV}}\end{array}\right).
\end{equation}
After traversing a HWP with $\hat{V}_{\mathrm{HWP}}(\phi)$, we have the new polarization coherence matrix:
\begin{equation}
\mathbf{G}'(\phi)=\left(\begin{array}{cc}\tfrac{1}{2}+\tfrac{\Delta}{2}C+G_{\mathrm{HV}}^{(\mathrm{R})}S&\tfrac{\Delta}{2}S-iG_{\mathrm{HV}}^{(\mathrm{I})}-G_{\mathrm{HV}}^{(\mathrm{R})}C\\\tfrac{\Delta}{2}S+iG_{\mathrm{HV}}^{(\mathrm{I})}-G_{\mathrm{HV}}^{(\mathrm{R})}C&\tfrac{1}{2}-\tfrac{\Delta}{2}C-G_{\mathrm{HV}}^{(\mathrm{R})}S\end{array}\right),
\end{equation}
where $G_{\mathrm{HV}}^{(\mathrm{R})}\!=\!\mathrm{Re}\{H_{\mathrm{HV}}\}$, $G_{\mathrm{HV}}^{(\mathrm{I})}\!=\!\mathrm{Im}\{H_{\mathrm{HV}}\}$, $C\!=\!\cos4\phi$, $S\!=\!\sin4\phi$, and $\Delta\!=\!G_{\mathrm{HH}}-G_{\mathrm{VV}}$. After integrating over $\phi$ (which we achieve by rotating the HWP and averaging the detected power over multiple full rotations), we have the polarization coherence matrix:
\begin{equation}
\mathbf{G}'=\tfrac{1}{2\pi}\int_{0}^{2\pi}\!\!d\phi\;\mathbf{G}'(\phi)=\left(\begin{array}{cc}\tfrac{1}{2}&-iG_{\mathrm{HV}}^{(\mathrm{I})}\\iG_{\mathrm{HV}}^{(\mathrm{I})}&\tfrac{1}{2}\end{array}\right).
\end{equation}
When the off-diagonal element $G_{\mathrm{HV}}$ is real, we have $\mathbf{G}'\!=\!\tfrac{1}{2}\mathbb{I}$ (fully unpolarized) independently of the initial state of polarization. In our experiments here, the off-diagonal elements are all zero, so this configuration suffices to act as a polarization randomizer. When placed at $a$, this device converts $\mathrm{G}\!=\!\mathrm{diag}\{\lambda_{1},\lambda_{2},\lambda_{3},\lambda_{4}\}$ to $\mathbf{G}'\!=\!\mathrm{diag}\{\tfrac{\lambda_{1}+\lambda_{2}}{2},\tfrac{\lambda_{1}+\lambda_{2}}{2},\lambda_{3},\lambda_{4}\}$, and when placed at $b$, it converts $\mathbf{G}$ to $\mathbf{G}'\!=\!\mathrm{diag}\{\lambda_{1},\lambda_{2},\tfrac{\lambda_{3}+\lambda_{4}}{2},\tfrac{\lambda_{3}+\lambda_{4}}{2}\}$.

\section{Synthesis of iso-entropy fields}

\noindent\textbf{Rank-2 fields:}

We synthesized rank-2 fields associated with the entropy values $S\!=\!0.5,0.75$, and 1.0.

\begin{figure}[t!]
\includegraphics[width=8.6cm]{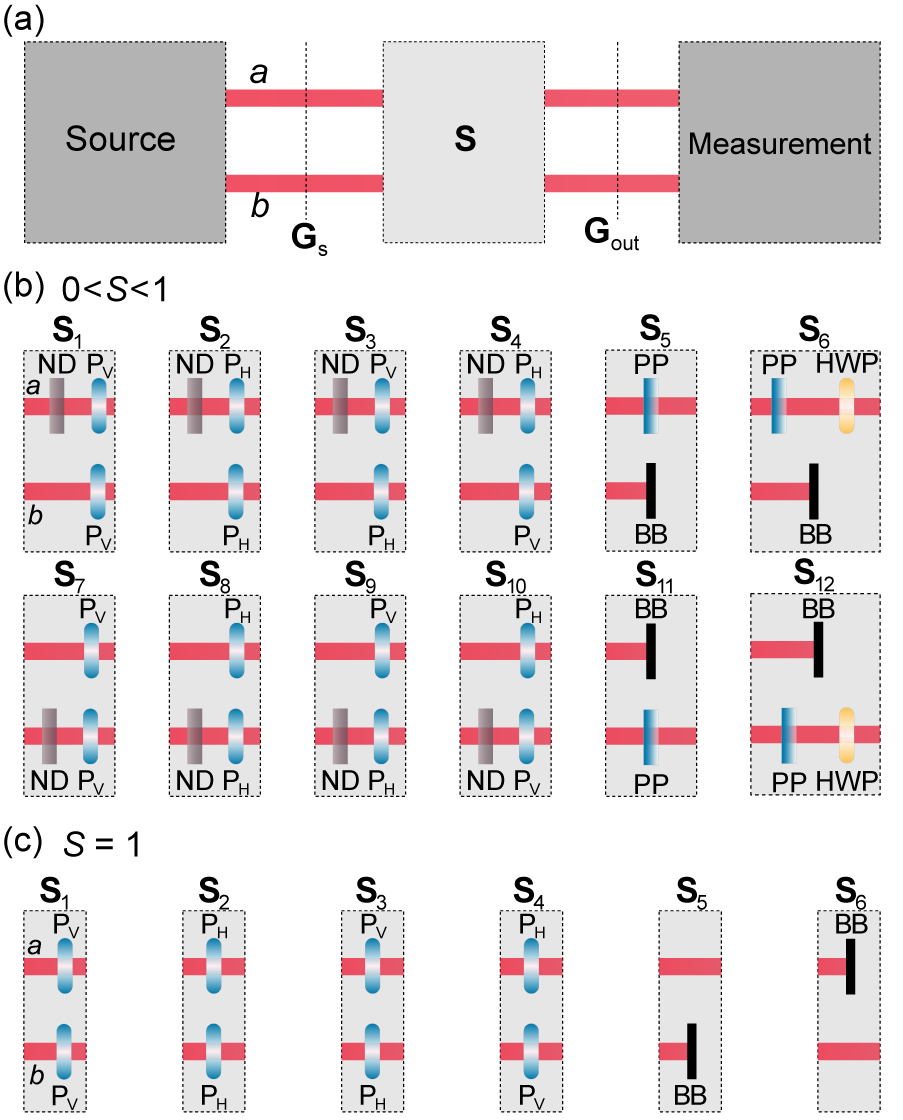}
\caption{Synthesizing rank-2 coherence matrices (a) Overall configuration used in the experiments. (b) The 12 transformations ($\mathbf{S}_{1}$ through $\mathbf{S}_{12}$) required to produce the permutations corresponding to the data plotted in Fig.~4(a,b) in the main text for $S\!=\!0.5$ and $S\!=\!0.75$. (c) The 6 transformations ($\mathbf{S}_{1}$ through $\mathbf{S}_{6}$) required to produce the permutations corresponding to the data plotted in Fig.~4(c) in the main text for $S\!=\!1$.}
\label{fig:rank2setups}
\end{figure}

\noindent\textit{Rank-2 with entropy of 0.5:} We selected the coherence matrix given theoretically by:
\begin{equation}
\mathbf{G}=\mathrm{diag}\{0.890, 0.110, 0, 0\}. \end{equation}
In our experiments, we produced optical fields represented by the following permutations of this coherence matrix:
\begin{eqnarray}
\mathbf{G}_{1}&=&\mathrm{diag}\{0.01, 0.10, 0.01, 0.88\},\nonumber\\
\mathbf{G}_{2}&=&\mathrm{diag}\{0.10, 0.01, 0.85, 0.03\},\nonumber\\
\mathbf{G}_{3}&=&\mathrm{diag}\{0.01, 0.11, 0.85, 0.03\},\nonumber\\
\mathbf{G}_{4}&=&\mathrm{diag}\{0.09, 0.01, 0.01, 0.89\},\nonumber\\
\mathbf{G}_{5}&=&\mathrm{diag}\{0.89, 0.11, 0, 0\},\nonumber\\
\mathbf{G}_{6}&=&\mathrm{diag}\{0.10,0.90, 0, 0\},\nonumber\\
\mathbf{G}_{7}&=&\mathrm{diag}\{0.02, 0.86, 0, 0.12\},\nonumber\\
\mathbf{G}_{8}&=&\mathrm{diag}\{0.86, 0.02, 0.11, 0.02\},\nonumber\\
\mathbf{G}_{9}&=&\mathrm{diag}\{0.02, 0.87, 0.10, 0.02\},\nonumber\\
\mathbf{G}_{10}&=&\mathrm{diag}\{0.85, 0.02, 0.00, 0.13\},\nonumber\\
\mathbf{G}_{11}&=&\mathrm{diag}\{0, 0, 0.88, 0.12\},\nonumber\\
\mathbf{G}_{12}&=&\mathrm{diag}\{0, 0, 0.10, 0.90\}.
\end{eqnarray}
These are the experimentally reconstructed coherence matrices using the overall scheme illustrated in Fig.~\ref{fig:rank2setups}(a). The data points representing these fields are plotted in Fig.~4(a) in the main text. The optical transformations used to synthesize these fields are depicted in Fig.~\ref{fig:rank2setups}(b), and are given by:
\begin{eqnarray}
\mathbf{S}_{1}&=&\left(\begin{array}{cc}
\hat{V}_{\mathrm{V}} \cdot \hat{V}_{\mathrm{ND}}\left(0.352\right)&\mathbf{0}\\
\mathbf{0}&\hat{V}_{\mathrm{V}}
\end{array}\right)\nonumber
,\\
\mathbf{S}_{2}&=&\left(\begin{array}{cc}
\hat{V}_{\mathrm{H}} \cdot \hat{V}_{\mathrm{ND}}\left(0.352\right)&\mathbf{0}\\
\mathbf{0}&\hat{V}_{\mathrm{H}}
\end{array}\right)\nonumber
,\\
\mathbf{S}_{3}&=&\left(\begin{array}{cc}
\hat{V}_{\mathrm{V}} \cdot \hat{V}_{\mathrm{ND}}\left(0.352\right)&\mathbf{0}\\
\mathbf{0}&\hat{V}_{\mathrm{H}}
\end{array}\right)\nonumber
,\\
\mathbf{S}_{4}&=&\left(\begin{array}{cc}
\hat{V}_{\mathrm{H}} \cdot \hat{V}_{\mathrm{ND}}\left(0.352\right)&\mathbf{0}\\
\mathbf{0}&\hat{V}_{\mathrm{V}}
\end{array}\right)\nonumber
,\\
\mathbf{S}_{5}&=&\left(\begin{array}{cc}
\hat{V}_{\mathrm{PP}}(0.943, 0.332)&\mathbf{0}\\
\mathbf{0}&\hat{V}_{\mathrm{BB}}
\end{array}\right)\nonumber
,\\
\mathbf{S}_{6}&=&\left(\begin{array}{cc}
\hat{V}_{\mathrm{HWP}}(45^\circ) \cdot \hat{V}_{\mathrm{PP}}(0.943, 0.332)&\mathbf{0}\\
\mathbf{0}&\hat{V}_{\mathrm{BB}}
\end{array}\right)\nonumber
,\\
\mathbf{S}_{7}&=&\left(\begin{array}{cc}
\hat{V}_{\mathrm{V}}&\mathbf{0}\\
\mathbf{0}&\hat{V}_{\mathrm{V}} \cdot \hat{V}_{\mathrm{ND}}\left(0.352\right)
\end{array}\right)\nonumber
,\\
\mathbf{S}_{8}&=&\left(\begin{array}{cc}
\hat{V}_{\mathrm{H}}&\mathbf{0}\\
\mathbf{0}&\hat{V}_{\mathrm{H}} \cdot \hat{V}_{\mathrm{ND}}\left(0.352\right)
\end{array}\right)\nonumber
,\\
\mathbf{S}_{9}&=&\left(\begin{array}{cc}
\hat{V}_{\mathrm{V}}&\mathbf{0}\\
\mathbf{0}&\hat{V}_{\mathrm{H}} \cdot \hat{V}_{\mathrm{ND}}\left(0.352\right)
\end{array}\right)\nonumber
,\\
\mathbf{S}_{10}&=&\left(\begin{array}{cc}
\hat{V}_{\mathrm{H}}&\mathbf{0}\\
\mathbf{0}&\hat{V}_{\mathrm{V}} \cdot \hat{V}_{\mathrm{ND}}\left(0.352\right)
\end{array}\right)\nonumber
,\\
\mathbf{S}_{11}&=&\left(\begin{array}{cc}
\hat{V}_{\mathrm{BB}}&\mathbf{0}\\
\mathbf{0}&\hat{V}_{\mathrm{PP}}(0.943, 0.332)
\end{array}\right)\nonumber
,\\
\mathbf{S}_{12}&=&\left(\begin{array}{cc}
\hat{V}_{\mathrm{BB}}&\mathbf{0}\\
\mathbf{0}&\hat{V}_{\mathrm{HWP}}(45^\circ) \cdot \hat{V}_{\mathrm{PP}}(0.943, 0.332)
\end{array}\right).
\label{eq: Rank-2 S=0.5 transformations}
\end{eqnarray}

\noindent\textit{Rank-2 with entropy of 0.75:} We selected the coherence matrix given theoretically by:
\begin{equation}
\mathbf{G}=\mathrm{diag}\{0.785, 0.215, 0, 0\},
\end{equation}
In our experiments, we produced optical fields represented by the following permutations of this coherence matrix:
\begin{eqnarray}
\mathbf{G}_{1}&=&\mathrm{diag}\{0.01, 0.20, 0.01, 0.77\},\nonumber\\
\mathbf{G}_{2}&=&\mathrm{diag}\{0.21, 0, 0.77, 0.02\},\nonumber\\
\mathbf{G}_{3}&=&\mathrm{diag}\{0.01, 0.20, 0.77, 0.02\},\nonumber\\
\mathbf{G}_{4}&=&\mathrm{diag}\{0.20, 0.01, 0.01, 0.77\},\nonumber\\
\mathbf{G}_{5}&=&\mathrm{diag}\{0.78,0.22,0,0\},\nonumber\\
\mathbf{G}_{6}&=&\mathrm{diag}\{0.19,0.81,0, 0\},\nonumber\\
\mathbf{G}_{7}&=&\mathrm{diag}\{0.01, 0.78, 0, 0.21\},\nonumber\\
\mathbf{G}_{8}&=&\mathrm{diag}\{0.79, 0, 0.21, 0\},\nonumber\\
\mathbf{G}_{9}&=&\mathrm{diag}\{0.01,0.78,0.21, 0.01\},\nonumber\\
\mathbf{G}_{10}&=&\mathrm{diag}\{0.79, 0, 0, 0.21\},\nonumber\\
\mathbf{G}_{11}&=&\mathrm{diag}\{0,0,0.79, 0.21\},\nonumber\\
\mathbf{G}_{12}&=&\mathrm{diag}\{0,0,0.18, 0.82\}.
\end{eqnarray}
These are the experimentally reconstructed coherence matrices. The data points representing these fields are plotted in Fig.~4(b) in the main text. The optical transformations used to synthesize these fields are depicted in Fig.~\ref{fig:rank2setups}(c), and are given by:
\begin{eqnarray}
\mathbf{S}_{1}&=&\left(\begin{array}{cc}
\hat{V}_{\mathrm{V}} \cdot \hat{V}_{\mathrm{ND}}\left(0.523\right)&\mathbf{0}\\
\mathbf{0}&\hat{V}_{\mathrm{V}}
\end{array}\right)\nonumber
,\\
\mathbf{S}_{2}&=&\left(\begin{array}{cc}
\hat{V}_{\mathrm{H}} \cdot \hat{V}_{\mathrm{ND}}\left(0.523\right)&\mathbf{0}\\
\mathbf{0}&\hat{V}_{\mathrm{H}}
\end{array}\right)\nonumber
,\\
\mathbf{S}_{3}&=&\left(\begin{array}{cc}
\hat{V}_{\mathrm{V}} \cdot \hat{V}_{\mathrm{ND}}\left(0.523\right)&\mathbf{0}\\
\mathbf{0}&\hat{V}_{\mathrm{H}}
\end{array}\right)\nonumber
,\\
\mathbf{S}_{4}&=&\left(\begin{array}{cc}
\hat{V}_{\mathrm{H}} \cdot \hat{V}_{\mathrm{ND}}\left(0.523\right)&\mathbf{0}\\
\mathbf{0}&\hat{V}_{\mathrm{V}}
\end{array}\right)\nonumber
,\\
\mathbf{S}_{5}&=&\left(\begin{array}{cc}
\hat{V}_{\mathrm{PP}}(0.886, 0.464)&\mathbf{0}\\
\mathbf{0}&\hat{V}_{\mathrm{BB}}
\end{array}\right)\nonumber
,\\
\mathbf{S}_{6}&=&\left(\begin{array}{cc}
\hat{V}_{\mathrm{HWP}}(45^\circ) \cdot \hat{V}_{\mathrm{PP}}(0.886, 0.464)&\mathbf{0}\\
\mathbf{0}&\hat{V}_{\mathrm{BB}}
\end{array}\right)\nonumber
,\\
\mathbf{S}_{7}&=&\left(\begin{array}{cc}
\hat{V}_{\mathrm{V}}&\mathbf{0}\\
\mathbf{0}&\hat{V}_{\mathrm{V}} \cdot \hat{V}_{\mathrm{ND}}\left(0.523\right)
\end{array}\right)\nonumber
,\\
\mathbf{S}_{8}&=&\left(\begin{array}{cc}
\hat{V}_{\mathrm{H}}&\mathbf{0}\\
\mathbf{0}&\hat{V}_{\mathrm{H}} \cdot \hat{V}_{\mathrm{ND}}\left(0.523\right)
\end{array}\right)\nonumber
,\\
\mathbf{S}_{9}&=&\left(\begin{array}{cc}
\hat{V}_{\mathrm{V}}&\mathbf{0}\\
\mathbf{0}&\hat{V}_{\mathrm{H}} \cdot \hat{V}_{\mathrm{ND}}\left(0.523\right)
\end{array}\right)\nonumber
,\\
\mathbf{S}_{10}&=&\left(\begin{array}{cc}
\hat{V}_{\mathrm{H}}&\mathbf{0}\\
\mathbf{0}&\hat{V}_{\mathrm{V}} \cdot \hat{V}_{\mathrm{ND}}\left(0.523\right)
\end{array}\right)\nonumber
,\\
\mathbf{S}_{11}&=&\left(\begin{array}{cc}
\hat{V}_{\mathrm{BB}}&\mathbf{0}\\
\mathbf{0}&\hat{V}_{\mathrm{PP}}(0.886, 0.464)
\end{array}\right)\nonumber
,\\
\mathbf{S}_{12}&=&\left(\begin{array}{cc}
\hat{V}_{\mathrm{BB}}&\mathbf{0}\\
\mathbf{0}&\hat{V}_{\mathrm{HWP}}(45^\circ) \cdot \hat{V}_{\mathrm{PP}}(0.886, 0.464)
\end{array}\right).
\label{eq: Rank-2 S=0.75 transformations}
\end{eqnarray}

\noindent\textit{Rank-2 with entropy of 1.0:} We selected the coherence matrix given theoretically by:
\begin{equation}
\mathbf{G}=\mathrm{diag}\{0.5,0.5, 0, 0\},
\end{equation}
In our experiments, we produced optical fields represented by the following permutations of this coherence matrix:
\begin{eqnarray}
\mathbf{G}_{1}&=&\mathrm{diag}\{0, 0.50, 0, 0.50\},\nonumber\\
\mathbf{G}_{2}&=&\mathrm{diag}\{0.49, 0, 0.50, 0\},\nonumber\\
\mathbf{G}_{3}&=&\mathrm{diag}\{0,0.52, 0.47, 0\},\nonumber\\
\mathbf{G}_{4}&=&\mathrm{diag}\{0.47, 0, 0, 0.53\},\nonumber\\
\mathbf{G}_{5}&=&\mathrm{diag}\{0.47,0.53, 0, 0\},\nonumber\\
\mathbf{G}_{6}&=&\mathrm{diag}\{0,0, 0.47, 0.53\}.
\end{eqnarray}
These are the experimentally reconstructed coherence matrices. The data points representing these fields are plotted in Fig.~4(c) in the main text. The optical transformations used to synthesize these fields are depicted in Fig.~\ref{fig:rank2setups}(c), and are given by: 
\begin{eqnarray}
\mathbf{S}_{1}&=&\left(\begin{array}{cc}
\hat{V}_{\mathrm{V}}&\mathbf{0}\\
\mathbf{0}&\hat{V}_{\mathrm{V}}
\end{array}\right),\nonumber\\
\mathbf{S}_{2}&=&\left(\begin{array}{cc}
\hat{V}_{\mathrm{H}}&\mathbf{0}\\
\mathbf{0}&\hat{V}_{\mathrm{H}}
\end{array}\right),\nonumber\\
\mathbf{S}_{3}&=&\left(\begin{array}{cc}
\hat{V}_{\mathrm{V}}&\mathbf{0}\\
\mathbf{0}&\hat{V}_{\mathrm{H}}
\end{array}\right),\nonumber\\
\mathbf{S}_{4}&=&\left(\begin{array}{cc}
\hat{V}_{\mathrm{H}}&\mathbf{0}\\
\mathbf{0}&\hat{V}_{\mathrm{V}}
\end{array}\right),\nonumber\\
\mathbf{S}_{5}&=&\left(\begin{array}{cc}
\mathbb{I}&\mathbf{0}\\
\mathbf{0}&\hat{V}_{\mathrm{BB}}
\end{array}\right),\nonumber\\
\mathbf{S}_{6}&=&\left(\begin{array}{cc}
\hat{V}_{\mathrm{BB}}&\mathbf{0}\\
\mathbf{0}&\mathbb{I}
\end{array}\right).
\label{eq: Rank-2 S=1 transformations}
\end{eqnarray}

\textbf{Rank-3 fields:}

We synthesized rank-3 fields associated with the entropy values $S\!=\!0.5, 0.75, 1.0, 1.25$, and 1.5.

\begin{figure}[h!]
\includegraphics[width=8.6cm]{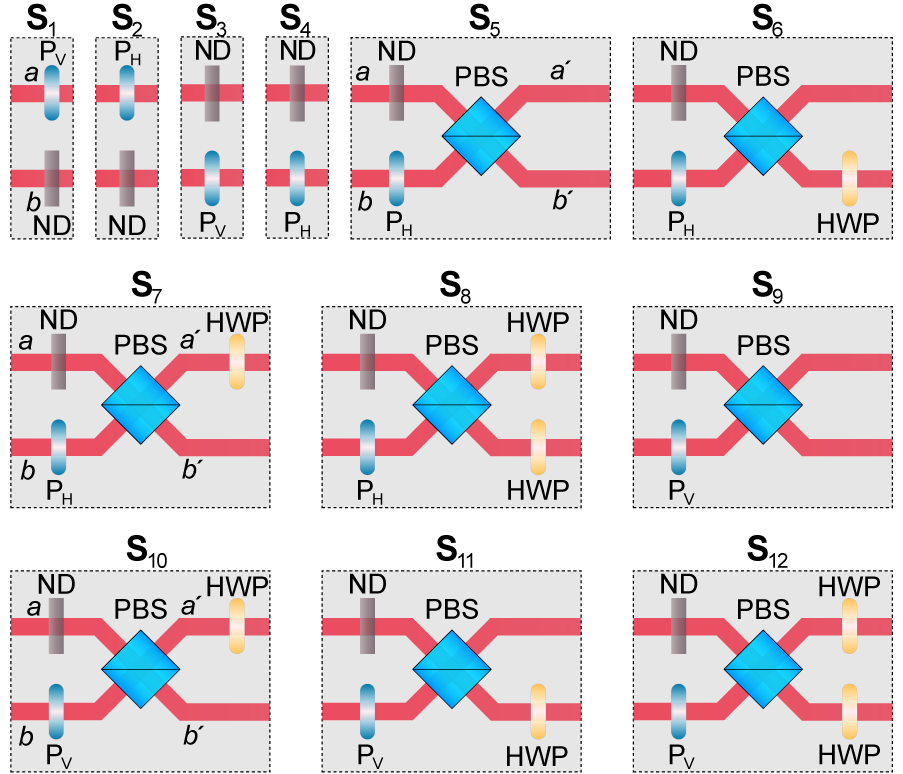}
\caption{Synthesizing rank-3 coherence matrices. The 12 transformations ($\mathbf{S}_{1}$ through $\mathbf{S}_{12}$) required to produce the permutations corresponding to the data plotted in Fig.~4(a,b) in the main text for $S\!=\!0.5$, $S\!=\!0.75$, and $S\!=\!1.0$.}
\label{fig:rank3setups}
\end{figure}

\noindent\textit{Rank-3 with entropy of 0.5:} We selected the coherence matrix given theoretically by:
\begin{equation}
\mathbf{G}=\mathrm{diag}\{0.916, 0.042, 0.042,0\}.
\end{equation}
In our experiments, we produced optical fields represented by the following permutations of this coherence matrix:
\begin{eqnarray}
\mathbf{G}_{1}&=&\mathrm{diag}\{0, 0.92, 0.03, 0.05\},\nonumber\\
\mathbf{G}_{2}&=&\mathrm{diag}\{0.90, 0.02, 0.04, 0.04\},\nonumber\\
\mathbf{G}_{3}&=&\mathrm{diag}\{0.03, 0.05, 0.01, 0.91\},\nonumber\\
\mathbf{G}_{4}&=&\mathrm{diag}\{0.03, 0.06, 0.90, 0.01\},\nonumber\\
\mathbf{G}_{5}&=&\mathrm{diag}\{0.92, 0.05, 0.02, 0.02\},\nonumber\\
\mathbf{G}_{6}&=&\mathrm{diag}\{0.93, 0.05, 0, 0.03\},\nonumber\\
\mathbf{G}_{7}&=&\mathrm{diag}\{0.04, 0.90, 0.05, 0.01\},\nonumber\\
\mathbf{G}_{8}&=&\mathrm{diag}\{0.04, 0.91, 0, 0.05\},\nonumber\\
\mathbf{G}_{9}&=&\mathrm{diag}\{0, 0.05, 0.04, 0.92\},\nonumber\\
\mathbf{G}_{10}&=&\mathrm{diag}\{0.03, 0.01, 0.04, 0.92\},\nonumber\\
\mathbf{G}_{11}&=&\mathrm{diag}\{0, 0.05, 0.91, 0.05\},\nonumber\\
\mathbf{G}_{12}&=&\mathrm{diag}\{0.03, 0.01, 0.91, 0.05\}.
\end{eqnarray}
These are the experimentally reconstructed coherence matrices. The data points representing these fields are plotted in Fig.~4(a) in the main text. The optical transformations used to synthesize these fields are depicted in Fig.~\ref{fig:rank3setups}(b), and are given by:
\begin{eqnarray}
\mathbf{S}_{1}&=&\left(\begin{array}{cc}
\hat{V}_{\mathrm{V}}&\mathbf{0}\\
\mathbf{0}&\hat{V}_{\mathrm{ND}}\left(0.214\right)
\end{array}\right),\nonumber\\
\mathbf{S}_{2}&=&\left(\begin{array}{cc}
\hat{V}_{\mathrm{H}}&\mathbf{0}\\
\mathbf{0}&\hat{V}_{\mathrm{ND}}\left(0.214\right)
\end{array}\right),\nonumber\\
\mathbf{S}_{3}&=&\left(\begin{array}{cc}
\hat{V}_{\mathrm{ND}}\left(0.214\right)&\mathbf{0}\\
\mathbf{0}&\hat{V}_{\mathrm{V}}
\end{array}\right),\nonumber\\
\mathbf{S}_{4}&=&\left(\begin{array}{cc}
\hat{V}_{\mathrm{ND}}\left(0.214\right)&\mathbf{0}\\
\mathbf{0}&\hat{V}_{\mathrm{H}}
\end{array}\right),\nonumber\\
\mathbf{S}_{5}&=&
\hat{V}_{\mathrm{PBS}} \cdot \mathbf{S}_4,\nonumber\\
\mathbf{S}_{6}&=&
\left(\begin{array}{cc}
\mathbb{I}&\mathbf{0}\\
\mathbf{0}&\hat{V}_{\mathrm{HWP}}(45^\circ)
\end{array}\right)
\cdot \mathbf{S}_5,\nonumber\\
\mathbf{S}_{7}&=&
\left(\begin{array}{cc}
\hat{V}_{\mathrm{HWP}}(45^\circ)&\mathbf{0}\\
\mathbf{0}&\mathbb{I}
\end{array}\right)
\cdot \mathbf{S}_5,\nonumber\\
\mathbf{S}_{8}&=&
\left(\begin{array}{cc}
\hat{V}_{\mathrm{HWP}}(45^\circ)&\mathbf{0}\\
\mathbf{0}&\hat{V}_{\mathrm{HWP}}(45^\circ)
\end{array}\right)
\cdot \mathbf{S}_5,\nonumber\\
\mathbf{S}_{9}&=&
\hat{V}_{\mathrm{PBS}} \cdot \mathbf{S}_3,\nonumber\\
\mathbf{S}_{10}&=&
\left(\begin{array}{cc}
\hat{V}_{\mathrm{HWP}}(45^\circ)&\mathbf{0}\\
\mathbf{0}&\mathbb{I}
\end{array}\right)
\cdot \mathbf{S}_9,\nonumber\\
\mathbf{S}_{11}&=&
\left(\begin{array}{cc}
\mathbb{I}&\mathbf{0}\\
\mathbf{0}&\hat{V}_{\mathrm{HWP}}(45^\circ)
\end{array}\right)
\cdot \mathbf{S}_9,\nonumber\\
\mathbf{S}_{12}&=&
\left(\begin{array}{cc}
\hat{V}_{\mathrm{HWP}}(45^\circ)&\mathbf{0}\\
\mathbf{0}&\hat{V}_{\mathrm{HWP}}(45^\circ)
\end{array}\right)
\cdot \mathbf{S}_9.
\label{eq: Rank-3 S=0.5 transformations}
\end{eqnarray}

\noindent\textit{Rank-3 with entropy of 0.75:} We selected the coherence matrix given theoretically by:
\begin{equation}
\mathbf{G}=\mathrm{diag}\{0.852, 0.074, 0.074,0\}.
\end{equation}
In our experiments, we produced optical fields represented by the following permutations of this coherence matrix:
\begin{eqnarray}
\mathbf{G}_{1}&=&\mathrm{diag}\{0, 0.85, 0.07, 0.08\},\nonumber\\
\mathbf{G}_{2}&=&\mathrm{diag}\{0.84, 0.02, 0.07, 0.08\},\nonumber\\
\mathbf{G}_{3}&=&\mathrm{diag}\{0.06, 0.08, 0.01, 0.85\},\nonumber\\
\mathbf{G}_{4}&=&\mathrm{diag}\{0.07, 0.08, 0.84, 0.01\},\nonumber\\
\mathbf{G}_{5}&=&\mathrm{diag}\{0.84, 0.08, 0.06, 0.2\},\nonumber\\
\mathbf{G}_{6}&=&\mathrm{diag}\{0.85, 0.08, 0, 0.07\},\nonumber\\
\mathbf{G}_{7}&=&\mathrm{diag}\{0.06, 0.87, 0.06, 0.02\},\nonumber\\
\mathbf{G}_{8}&=&\mathrm{diag}\{0.06, 0.87, 0, 0.07\},\nonumber\\
\mathbf{G}_{9}&=&\mathrm{diag}\{0, 0.07, 0.07, 0.86\},\nonumber\\
\mathbf{G}_{10}&=&\mathrm{diag}\{0.06, 0.01, 0.07, 0.86\},\nonumber\\
\mathbf{G}_{11}&=&\mathrm{diag}\{0, 0.08, 0.84, 0.09\},\nonumber\\
\mathbf{G}_{12}&=&\mathrm{diag}\{0.06, 0.01, 0.84, 0.09\}.
\end{eqnarray}
These are the experimentally reconstructed coherence matrices. The data points representing these fields are plotted in Fig.~4(b) in the main text. The optical transformations used to synthesize these fields are depicted in Fig.~\ref{fig:rank3setups}, and are given by:
\begin{eqnarray}
\mathbf{S}_{1}&=&\left(\begin{array}{cc}
\hat{V}_{\mathrm{V}}&\mathbf{0}\\
\mathbf{0}&\hat{V}_{\mathrm{ND}}(0.295)
\end{array}\right),\nonumber\\
\mathbf{S}_{2}&=&\left(\begin{array}{cc}
\hat{V}_{\mathrm{H}}&\mathbf{0}\\
\mathbf{0}\hat{V}_{\mathrm{ND}}(0.295)
\end{array}\right),\nonumber\\
\mathbf{S}_{3}&=&\left(\begin{array}{cc}
\hat{V}_{\mathrm{ND}}(0.295)&\mathbf{0}\\
\mathbf{0}&\hat{V}_{\mathrm{V}}
\end{array}\right),\nonumber\\
\mathbf{S}_{4}&=&\left(\begin{array}{cc}
\hat{V}_{\mathrm{ND}}(0.295)&\mathbf{0}\\
\mathbf{0}&\hat{V}_{\mathrm{H}}
\end{array}\right),\nonumber\\
\mathbf{S}_{5}&=&\hat{V}_{\mathrm{PBS}} \cdot \mathbf{S}_{4},\nonumber\\
\mathbf{S}_{6}&=&\left(\begin{array}{cc}
\mathbb{I}&\mathbf{0}\\
\mathbf{0}&\hat{V}_{\mathrm{HWP}}(45^\circ)
\end{array}\right)\cdot\mathbf{S}_{5},\nonumber\\
\mathbf{S}_{7}&=&\left(\begin{array}{cc}
\hat{V}_{\mathrm{HWP}}(45^\circ)&\mathbf{0}\\
\mathbf{0}&\mathbb{I}\end{array}\right)
\cdot \mathbf{S}_{5},\nonumber\\
\mathbf{S}_{8}&=&\left(\begin{array}{cc}
\hat{V}_{\mathrm{HWP}}(45^\circ)&\mathbf{0}\\
\mathbf{0}&\hat{V}_{\mathrm{HWP}}(45^\circ)
\end{array}\right)
\cdot \mathbf{S}_{5},\nonumber\\
\mathbf{S}_{9}&=&\hat{V}_{\mathrm{PBS}} \cdot \mathbf{S}_{3},\nonumber\\
\mathbf{S}_{10}&=&\left(\begin{array}{cc}
\hat{V}_{\mathrm{HWP}}(45^\circ)&\mathbf{0}\\
\mathbf{0}\mathbb{I}\end{array}\right)
\cdot \mathbf{S}_{9},\nonumber\\
\mathbf{S}_{11}&=&\left(\begin{array}{cc}
\mathbb{I}&\mathbf{0}\\
\mathbf{0}&\hat{V}_{\mathrm{HWP}}(45^\circ)
\end{array}\right)\cdot\mathbf{S}_{9},\nonumber\\
\mathbf{S}_{12}&=&\left(\begin{array}{cc}
\hat{V}_{\mathrm{HWP}}(45^\circ)&\mathbf{0}\\
\mathbf{0}&\hat{V}_{\mathrm{HWP}}(45^\circ)
\end{array}\right)\cdot\mathbf{S}_{9}.
\label{eq: Rank-3 S=0.75 transformations}
\end{eqnarray}

\noindent\textit{Rank-3 with entropy of 1.0:} We selected the coherence matrix given theoretically by:
\begin{equation}
\mathbf{G}=\mathrm{diag}\{0.772, 0.114, 0.114, 0\}.
\end{equation}
In our experiments, we produced optical fields represented by the following permutations of this coherence matrix:
\begin{eqnarray}
\mathbf{G}_{1}&=&\mathrm{diag}\{0.01, 0.76, 0.10, 0.12\},\nonumber\\
\mathbf{G}_{2}&=&\mathrm{diag}\{0.77, 0, 0.10, 0.12\},\nonumber\\
\mathbf{G}_{3}&=&\mathrm{diag}\{0.10, 0.12, 0.01, 0.77\},\nonumber\\
\mathbf{G}_{4}&=&\mathrm{diag}\{0.10, 0.12, 0.77, 0.01\},\nonumber\\
\mathbf{G}_{5}&=&\mathrm{diag}\{0.77, 0.12, 0.10, 0.01\},\nonumber\\
\mathbf{G}_{6}&=&\mathrm{diag}\{0.10, 0.78, 0.10, 0.01\},\nonumber\\
\mathbf{G}_{7}&=&\mathrm{diag}\{0.77, 0.12, 0, 0.11\},\nonumber\\
\mathbf{G}_{8}&=&\mathrm{diag}\{0.10, 0.79, 0, 0.11\},\nonumber\\
\mathbf{G}_{9}&=&\mathrm{diag}\{0, 0.11, 0.11, 0.78\},\nonumber\\
\mathbf{G}_{10}&=&\mathrm{diag}\{0.11, 0.01, 0.11, 0.78\},\nonumber\\
\mathbf{G}_{11}&=&\mathrm{diag}\{0, 0.12, 0.77, 0.12\},\nonumber\\
\mathbf{G}_{12}&=&\mathrm{diag}\{0.11, 0.01, 0.77, 0.12\}.
\end{eqnarray}
These are the experimentally reconstructed coherence matrices. The data points representing these fields are plotted in Fig.~4(c) in the main text. The optical transformations used to synthesize these fields are depicted in Fig.~\ref{fig:rank3setups}, and are given by
\begin{eqnarray}
\mathbf{S}_{1}&=&\left(\begin{array}{cc}
\hat{V}_{\mathrm{V}}&\mathbf{0}\\
\mathbf{0}&\hat{V}_{\mathrm{ND}}\left(0.384\right)
\end{array}\right),\nonumber\\
\mathbf{S}_{2}&=&\left(\begin{array}{cc}
\hat{V}_{\mathrm{H}}&\mathbf{0}\\
\mathbf{0}&\hat{V}_{\mathrm{ND}}\left(0.384\right)
\end{array}\right),\nonumber\\
\mathbf{S}_{3}&=&\left(\begin{array}{cc}
\hat{V}_{\mathrm{ND}}\left(0.384\right)&\mathbf{0}\\
\mathbf{0}&\hat{V}_{\mathrm{V}}
\end{array}\right),\nonumber\\
\mathbf{S}_{4}&=&\left(\begin{array}{cc}
\hat{V}_{\mathrm{ND}}\left(0.384\right)&\mathbf{0}\\
\mathbf{0}&\hat{V}_{\mathrm{H}}
\end{array}\right),\nonumber\\
\mathbf{S}_{5}&=&\hat{V}_{\mathrm{PBS}} \cdot \mathbf{S}_{4},\nonumber\\
\mathbf{S}_{6}&=&\left(\begin{array}{cc}
\mathbb{I}&\mathbf{0}\\\mathbf{0}&\hat{V}_{\mathrm{HWP}}(45^\circ)\end{array}\right)
\cdot \mathbf{S}_{5},\nonumber\\
\mathbf{S}_{7}&=&\left(\begin{array}{cc}
\hat{V}_{\mathrm{HWP}}(45^\circ)&\mathbf{0}\\
\mathbf{0}&\mathbb{I}\end{array}\right)
\cdot \mathbf{S}_{5},\nonumber\\
\mathbf{S}_{8}&=&\left(\begin{array}{cc}
\hat{V}_{\mathrm{HWP}}(45^\circ)&\mathbf{0}\\
\mathbf{0}&\hat{V}_{\mathrm{HWP}}(45^\circ)
\end{array}\right)\cdot\mathbf{S}_{5},\nonumber\\
\mathbf{S}_{9}&=&\hat{V}_{\mathrm{PBS}} \cdot \mathbf{S}_{3},\nonumber\\
\mathbf{S}_{10}&=&\left(\begin{array}{cc}
\hat{V}_{\mathrm{HWP}}(45^\circ)&\mathbf{0}\\
\mathbf{0}&\mathbb{I}\end{array}\right)
\cdot \mathbf{S}_{9},\nonumber\\
\mathbf{S}_{11}&=&\left(\begin{array}{cc}
\mathbb{I}&\mathbf{0}\\
\mathbf{0}&\hat{V}_{\mathrm{HWP}}(45^\circ)
\end{array}\right)\cdot\mathbf{S}_{9},\nonumber\\
\mathbf{S}_{12}&=&\left(\begin{array}{cc}
\hat{V}_{\mathrm{HWP}}(45^\circ)&\mathbf{0}\\
\mathbf{0}&\hat{V}_{\mathrm{HWP}}(45^\circ)
\end{array}\right)\cdot\mathbf{S}_{9}.
\label{eq: Rank-3 S=1 transformations}
\end{eqnarray}

\noindent\textit{Rank-3 with entropy of 1.25:} We selected the coherence matrix given theoretically by:
\begin{equation}
\mathbf{G}=\mathrm{diag}\{0.668, 0.166, 0.166, 0\}.
\end{equation}
In our experiments, we produced optical fields represented by the following permutations of this coherence matrix:
\begin{eqnarray}
\mathbf{G}_{1}&=&\mathrm{diag}\{0.01, 0.67, 0.15, 0.17\},\nonumber\\
\mathbf{G}_{2}&=&\mathrm{diag}\{0.67, 0, 0.15, 0.18\},\nonumber\\
\mathbf{G}_{3}&=&\mathrm{diag}\{0.15, 0.18, 0.01, 0.66\},\nonumber\\
\mathbf{G}_{4}&=&\mathrm{diag}\{0.15, 0.18, 0.67, 0\},\nonumber\\
\mathbf{G}_{5}&=&\mathrm{diag}\{0.66, 0.17, 0.15, 0.02\},\nonumber\\
\mathbf{G}_{6}&=&\mathrm{diag}\{0.67, 0.17, 0, 0.16\},\nonumber\\
\mathbf{G}_{7}&=&\mathrm{diag}\{0.15, 0.69, 0.15, 0.02\},\nonumber\\
\mathbf{G}_{8}&=&\mathrm{diag}\{0.15, 0.69, 0, 0.16\},\nonumber\\
\mathbf{G}_{9}&=&\mathrm{diag}\{0.01, 0.17, 0.16, 0.67\},\nonumber\\
\mathbf{G}_{10}&=&\mathrm{diag}\{0.13, 0.01, 0.16, 0.70\},\nonumber\\
\mathbf{G}_{11}&=&\mathrm{diag}\{0.07, 0.18, 0.67, 0.14\},\nonumber\\
\mathbf{G}_{12}&=&\mathrm{diag}\{0.13, 0.01, 0.71, 0.15\}.
\end{eqnarray}
These are the experimentally reconstructed coherence matrices. The data points representing these fields are plotted in Fig.~4(d) in the main text. The optical transformations used to synthesize these fields are depicted in Fig.~\ref{fig:rank3setups}, and are given by:
\begin{eqnarray}
\mathbf{S}_{1}&=&\left(\begin{array}{cc}
\hat{V}_{\mathrm{V}}&\mathbf{0}\\
\mathbf{0}&\hat{V}_{\mathrm{ND}}\left(0.417\right)
\end{array}\right),\nonumber\\
\mathbf{S}_{2}&=&\left(\begin{array}{cc}
\hat{V}_{\mathrm{H}}&\mathbf{0}\\
\mathbf{0}&\hat{V}_{\mathrm{ND}}\left(0.417\right)
\end{array}\right),\nonumber\\
\mathbf{S}_{3}&=&\left(\begin{array}{cc}
\hat{V}_{\mathrm{ND}}\left(0.417\right)&\mathbf{0}\\
\mathbf{0}&\hat{V}_{\mathrm{V}}
\end{array}\right),\nonumber\\
\mathbf{S}_{4}&=&\left(\begin{array}{cc}
\hat{V}_{\mathrm{ND}}\left(0.417\right)&\mathbf{0}\\
\mathbf{0}&\hat{V}_{\mathrm{H}}
\end{array}\right),\nonumber\\
\mathbf{S}_{5}&=&\hat{V}_{\mathrm{PBS}} \cdot \mathbf{S}_{4},\nonumber\\
\mathbf{S}_{6}&=&\left(\begin{array}{cc}
\mathbb{I}&\mathbf{0}\\
\mathbf{0}&\hat{V}_{\mathrm{HWP}}(45^\circ)
\end{array}\right)\cdot\mathbf{S}_{5},\nonumber\\
\mathbf{S}_{7}&=&\left(\begin{array}{cc}
\hat{V}_{\mathrm{HWP}}(45^\circ)&\mathbf{0}\\
\mathbf{0}&\mathbb{I}\end{array}\right)
\cdot \mathbf{S}_{5},\nonumber\\
\mathbf{S}_{8}&=&\left(\begin{array}{cc}
\hat{V}_{\mathrm{HWP}}(45^\circ)&\mathbf{0}\\
\mathbf{0}&\hat{V}_{\mathrm{HWP}}(45^\circ)
\end{array}\right)\cdot\mathbf{S}_{5},\nonumber\\
\mathbf{S}_{9}&=&\hat{V}_{\mathrm{PBS}} \cdot \mathbf{S}_{3},\nonumber\\
\mathbf{S}_{10}&=&\left(\begin{array}{cc}
\hat{V}_{\mathrm{HWP}}(45^\circ)&\mathbf{0}\\
\mathbf{0}&\mathbb{I}\end{array}\right)
\cdot \mathbf{S}_{9},\nonumber\\
\mathbf{S}_{11}&=&\left(\begin{array}{cc}
\mathbb{I}&\mathbf{0}\\
\mathbf{0}&\hat{V}_{\mathrm{HWP}}(45^\circ)
\end{array}\right)\cdot\mathbf{S}_{9},\nonumber\\
\mathbf{S}_{12}&=&\left(\begin{array}{cc}
\hat{V}_{\mathrm{HWP}}(45^\circ)&\mathbf{0}\\
\mathbf{0}&\hat{V}_{\mathrm{HWP}}(45^\circ)
\end{array}\right)\cdot\mathbf{S}_{9}.
\label{eq: Rank-3 S=1.25 transformations}
\end{eqnarray}

\noindent\textit{Rank-3 with entropy of 1.5.} We selected the coherence matrix given theoretically by:
\begin{equation}
\mathbf{G}=\mathrm{diag}\{0.5, 0.25, 0.25, 0\}.
\end{equation}
In our experiments, we produced optical fields represented by the following permutations of this coherence matrix:
\begin{eqnarray}
\mathbf{G}_{1}&=&\mathrm{diag}\{0, 0.51, 0.23, 0.26\},\nonumber\\
\mathbf{G}_{2}&=&\mathrm{diag}\{0.51, 0.01, 0.23, 0.26\},\nonumber\\
\mathbf{G}_{3}&=&\mathrm{diag}\{0.24, 0.27, 0, 0.49\},\nonumber\\
\mathbf{G}_{4}&=&\mathrm{diag}\{0.24, 0.27, 0.48, 0.01\},\nonumber\\
\mathbf{G}_{5}&=&\mathrm{diag}\{0, 0.27, 0.21, 0.52\},\nonumber\\
\mathbf{G}_{6}&=&\mathrm{diag}\{0, 0.27, 0.51, 0.22\},\nonumber\\
\mathbf{G}_{7}&=&\mathrm{diag}\{0.24, 0.01, 0.22, 0.53\},\nonumber\\
\mathbf{G}_{8}&=&\mathrm{diag}\{0.24, 0.01, 0.52, 0.23\},\nonumber\\
\mathbf{G}_{9}&=&\mathrm{diag}\{0.49, 0.28, 0.22, 0.01\},\nonumber\\
\mathbf{G}_{10}&=&\mathrm{diag}\{0.26, 0.50, 0.23, 0.01\},\nonumber\\
\mathbf{G}_{11}&=&\mathrm{diag}\{0.49, 0.28, 0, 0.23\},\nonumber\\
\mathbf{G}_{12}&=&\mathrm{diag}\{0.26, 0.50, 0, 0.24\}
\end{eqnarray}
These are the experimentally reconstructed coherence matrices. The data points representing these fields are plotted in Fig.~4(e) in the main text. The optical transformations used to synthesize these fields are depicted in Fig.~\ref{fig:rank3setups}(b), and are given by:
\begin{eqnarray}
\mathbf{S}_{1}&=&\left(\begin{array}{cc}
\hat{V}_{\mathrm{V}}&\mathbf{0}\\
\mathbf{0}&\hat{V}_{\mathrm{ND}}
(0.707)\end{array}\right),\nonumber\\
\mathbf{S}_{2}&=&\left(\begin{array}{cc}
\hat{V}_{\mathrm{H}}&\mathbf{0}\\
\mathbf{0}&\hat{V}_{\mathrm{ND}}(0.707)
\end{array}\right),\nonumber\\
\mathbf{S}_{3}&=&\left(\begin{array}{cc}
\hat{V}_{\mathrm{ND}}(0.707)&\mathbf{0}\\
\mathbf{0}&\hat{V}_{\mathrm{V}}
\end{array}\right),\nonumber\\
\mathbf{S}_{4}&=&\left(\begin{array}{cc}
\hat{V}_{\mathrm{ND}}(0.707)&\mathbf{0}\\
\mathbf{0}&\hat{V}_{\mathrm{H}}
\end{array}\right),\nonumber\\
\mathbf{S}_{5}&=&\hat{V}_{\mathrm{PBS}} \cdot \mathbf{S}_{4},\nonumber\\
\mathbf{S}_{6}&=&\left(\begin{array}{cc}
\mathbb{I}&\mathbf{0}\\
\mathbf{0}&\hat{V}_{\mathrm{HWP}}(45^\circ)
\end{array}\right)\cdot\mathbf{S}_{5},\nonumber\\
\mathbf{S}_{7}&=&\left(\begin{array}{cc}
\hat{V}_{\mathrm{HWP}}(45^\circ)&\mathbf{0}\\
\mathbf{0}&\mathbb{I}\end{array}\right)
\cdot \mathbf{S}_{5},\nonumber\\
\mathbf{S}_{8}&=&\left(\begin{array}{cc}
\hat{V}_{\mathrm{HWP}}(45^\circ)&\mathbf{0}\\
\mathbf{0}&\hat{V}_{\mathrm{HWP}}(45^\circ)
\end{array}\right)\cdot\mathbf{S}_{5},\nonumber\\
\mathbf{S}_{9}&=&\hat{V}_{\mathrm{PBS}} \cdot \mathbf{S}_{3},\nonumber\\
\mathbf{S}_{10}&=&\left(\begin{array}{cc}
\hat{V}_{\mathrm{HWP}}(45^\circ)&\mathbf{0}\\
\mathbf{0}&\mathbb{I}
\end{array}\right)\cdot\mathbf{S}_{9},\nonumber\\
\mathbf{S}_{11}&=&\left(\begin{array}{cc}
\mathbb{I}&\mathbf{0}\\
\mathbf{0}&\hat{V}_{\mathrm{HWP}}(45^\circ)
\end{array}\right)\cdot\mathbf{S}_{9},\nonumber\\
\mathbf{S}_{12}&=&\left(\begin{array}{cc}
\hat{V}_{\mathrm{HWP}}(45^\circ)&\mathbf{0}\\
\mathbf{0}&\hat{V}_{\mathrm{HWP}}(45^\circ)
\end{array}\right)\cdot\mathbf{S}_{9}.
\label{eq: Rank-3 S=1.25 transformations}
\end{eqnarray}

\textbf{Rank-4 fields}

\begin{figure}[t!]
\includegraphics[width=8.6cm]{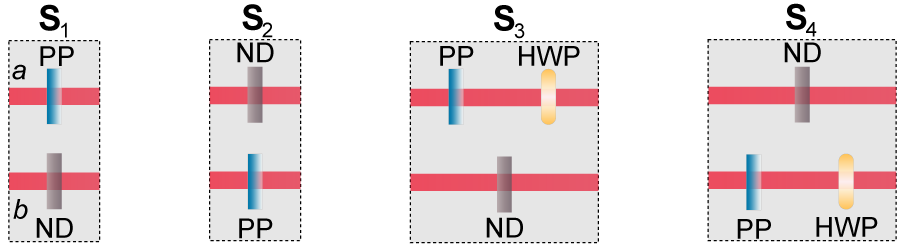}
\caption{Synthesizing rank-4 coherence matrices. The 4 transformations ($\mathbf{S}_{1}$ through $\mathbf{S}_{4}$) required to produce the permutations corresponding to the data plotted in Fig.~4(a-f).}
\label{fig:rank4setups}
\end{figure}

\noindent\textit{Rank-4 with entropy of 0.5:} We selected the coherence matrix given theoretically by:
\begin{equation}
\mathbf{G}=\mathrm{diag}\{0.925,0.025, 0.025, 0.025\}.
\end{equation}
In our experiments, we produced optical fields represented by the following permutations of this coherence matrix:
\begin{eqnarray}
\mathbf{G}_{1}&=&\mathrm{diag}\{0.90, 0.04, 0.02, 0.04\},\nonumber\\
\mathbf{G}_{2}&=&\mathrm{diag}\{0.02, 0.04, 0.90, 0.04\},\nonumber\\
\mathbf{G}_{3}&=&\mathrm{diag}\{0.03, 0.91, 0.02, 0.04\},\nonumber\\
\mathbf{G}_{4}&=&\mathrm{diag}\{0.02, 0.04, 0.02, 0.91\}.
\end{eqnarray}
These are the experimentally reconstructed coherence matrices. The data points representing these fields are plotted in Fig.~4(a) in the main text. The optical transformations used to synthesize these fields are depicted in Fig.~\ref{fig:rank4setups} are given by:
\begin{eqnarray}
\mathbf{S}_{1}&=&\left(\begin{array}{cc}
\hat{V}_{\mathrm{PP}}(0.925,0.025)&\mathbf{0}\\
\mathbf{0}&\hat{V}_{\mathrm{ND}}(0.025)
\end{array}\right),\nonumber\\
\mathbf{S}_{2}&=&\left(\begin{array}{cc}
\hat{V}_{\mathrm{ND}}(0.025)&\mathbf{0}\\
\mathbf{0}&\hat{V}_{\mathrm{PP}}(0.925,0.025)
\end{array}\right),\nonumber\\
\mathbf{S}_{3}&=&\left(\begin{array}{cc}
\hat{V}_{\mathrm{HWP}}(45^{\circ})&\mathbf{0}\\
\mathbf{0}&\mathbb{I}\end{array}\right)\cdot \mathbf{S}_{1},\nonumber\\
\mathbf{S}_{4}&=&\left(\begin{array}{cc}
\mathbb{I}&\mathbf{0}\\
\mathbf{0}&\hat{V}_{\mathrm{HWP}}(45^{\circ})
\end{array}\right)\cdot\mathbf{S}_{2}.
\label{eq: Rank-4 S=0.5 transformations}
\end{eqnarray}

\noindent\textit{Rank-4 with entropy of 0.75.} We selected the coherence matrix given theoretically by:
\begin{equation}
\mathbf{G}=\mathrm{diag}\{0.873,0.042, 0.042, 0.042\}.
\end{equation}
In our experiments, we produced optical fields represented by the following permutations of this coherence matrix:
\begin{eqnarray}
\mathbf{G}_{1}&=&\mathrm{diag}\{0.87, 0.05, 0.03, 0.04\},\nonumber\\
\mathbf{G}_{2}&=&\mathrm{diag}\{0.04, 0.05, 0.87, 0.05\},\nonumber\\
\mathbf{G}_{3}&=&\mathrm{diag}\{0.04, 0.88, 0.04, 0.05\},\nonumber\\
\mathbf{G}_{4}&=&\mathrm{diag}\{0.04, 0.05, 0.04, 0.88\}.
\end{eqnarray}
These are the experimentally reconstructed coherence matrices. The data points representing these fields are plotted in Fig.~4(b) in the main text. The optical transformations used to synthesize these fields are depicted in Fig.~\ref{fig:rank4setups} and are given by:
\begin{eqnarray}
\mathbf{S}_{1}&=&\left(\begin{array}{cc}
\hat{V}_{\mathrm{PP}}(0.811,0.063)&\mathbf{0}\\
\mathbf{0}&\hat{V}_{\mathrm{ND}}(0.063)
\end{array}\right),\nonumber\\
\mathbf{S}_{2}&=&\left(\begin{array}{cc}
\hat{V}_{\mathrm{ND}}(0.063)&\mathbf{0}\\
\mathbf{0}&\hat{V}_{\mathrm{PP}}(0.811,0.063)
\end{array}\right),\nonumber\\
\mathbf{S}_{3}&=&\left(\begin{array}{cc}
\hat{V}_{\mathrm{HWP}}(45^{\circ})&\mathbf{0}\\
\mathbf{0}&\mathbb{I}\end{array}\right)\cdot \mathbf{S}_{1},\nonumber\\
\mathbf{S}_{4}&=&\left(\begin{array}{cc}
\mathbb{I}&\mathbf{0}\\
\mathbf{0}&\hat{V}_{\mathrm{HWP}}(45^{\circ})
\end{array}\right)\cdot\mathbf{S}_{2}.
\label{eq: Rank-4 S=0.75 transformations}
\end{eqnarray}

\noindent\textit{Rank-4 with entropy of 1.0:} We selected the coherence matrix given theoretically by:
\begin{equation}
\mathbf{G}=\mathrm{diag}\{0.811,0.063, 0.063, 0.063\}.
\end{equation}
In our experiments, we produced optical fields represented by the following permutations of this coherence matrix:
\begin{eqnarray}
\mathbf{G}_{1}&=&\mathrm{diag}\{0.81, 0.07, 0.05, 0.07\},\nonumber\\
\mathbf{G}_{2}&=&\mathrm{diag}\{0.06, 0.07, 0.80, 0.07\},\nonumber\\
\mathbf{G}_{3}&=&\mathrm{diag}\{0.06, 0.81, 0.05, 0.08\},\nonumber\\
\mathbf{G}_{4}&=&\mathrm{diag}\{0.06, 0.08, 0.07, 0.80\}.
\end{eqnarray}
These are the experimentally reconstructed coherence matrices. The data points representing these fields are plotted in Fig.~4(c) in the main text. The optical transformations used to synthesize these fields are depicted in Fig.~\ref{fig:rank4setups} and are given by:
\begin{eqnarray}
\mathbf{S}_{1}&=&\left(\begin{array}{cc}
\hat{V}_{\mathrm{PP}}(0.811,0.063)&\mathbf{0}\\
\mathbf{0}&\hat{V}_{\mathrm{ND}}(0.063)
\end{array}\right),\nonumber\\
\mathbf{S}_{2}&=&\left(\begin{array}{cc}
\hat{V}_{\mathrm{ND}}(0.063)&\mathbf{0}\\
\mathbf{0}&\hat{V}_{\mathrm{PP}}(0.811,0.063)
\end{array}\right),\nonumber\\
\mathbf{S}_{3}&=&\left(\begin{array}{cc}
\hat{V}_{\mathrm{HWP}}(45^{\circ})&\mathbf{0}\\
\mathbf{0}&\mathbb{I}
\end{array}\right)\cdot\mathbf{S}_{1},\nonumber\\
\mathbf{S}_{4}&=&\left(\begin{array}{cc}
\mathbb{I}&\mathbf{0}\\
\mathbf{0}&\hat{V}_{\mathrm{HWP}}(45^{\circ})
\end{array}\right)\cdot\mathbf{S}_{2}.
\label{eq: Rank-4 S=1 transformations}
\end{eqnarray}

\noindent\textit{Rank-4 with entropy of 1.25:} We selected the coherence matrix given theoretically by:
\begin{equation}
\mathbf{G}=\mathrm{diag}\{0.736, 0.088, 0.088, 0.088\}.
\end{equation}
In our experiments, we produced optical fields represented by the following permutations of this coherence matrix:
\begin{eqnarray}
\mathbf{G}_{1}&=&\mathrm{diag}\{0.73, 0.09, 0.08, 0.10\},\nonumber\\
\mathbf{G}_{2}&=&\mathrm{diag}\{0.08, 0.11, 0.72, 0.09\},\nonumber\\
\mathbf{G}_{3}&=&\mathrm{diag}\{0.08, 0.73, 0.09, 0.10\},\nonumber\\
\mathbf{G}_{4}&=&\mathrm{diag}\{0.08, 0.11, 0.07, 0.74\}.
\end{eqnarray}
These are the experimentally reconstructed coherence matrices. The data points representing these fields are plotted in Fig.~4(d) in the main text. The optical transformations used to synthesize these fields are depicted in Fig.~\ref{fig:rank4setups} and are given by:
\begin{eqnarray}
\mathbf{S}_{1}&=&\left(\begin{array}{cc}
\hat{V}_{\mathrm{PP}}(0.736,0.088)&\mathbf{0}\\
\mathbf{0}&\hat{V}_{\mathrm{ND}}(0.088)
\end{array}\right),\nonumber\\
\mathbf{S}_{2}&=&\left(\begin{array}{cc}
\hat{V}_{\mathrm{ND}}(0.088)&\mathbf{0}\\
\mathbf{0}&\hat{V}_{\mathrm{PP}}(0.736,0.088)
\end{array}\right),\nonumber\\
\mathbf{S}_{3}&=&\left(\begin{array}{cc}
\hat{V}_{\mathrm{HWP}}(45^{\circ})&\mathbf{0}\\
\mathbf{0}&\mathbb{I}\end{array}\right)\cdot \mathbf{S}_{1},\nonumber\\
\mathbf{S}_{4}&=&\left(\begin{array}{cc}
\mathbb{I}&\mathbf{0}\\
\mathbf{0}&\hat{V}_{\mathrm{HWP}}(45^{\circ})
\end{array}\right) \cdot \mathbf{S}_{2}.
\label{eq: Rank-4 S=1.25 transformations}
\end{eqnarray}

\noindent\textit{Rank-4 with entropy of 1.5:} We selected the coherence matrix given theoretically by:
\begin{equation}
\mathbf{G}=\mathrm{diag}\{0.646, 0.118, 0.118, 0.118\},
\end{equation}
In our experiments, we produced optical fields represented by the following permutations of this coherence matrix:
\begin{eqnarray}
\mathbf{G}_{1}&=&\mathrm{diag}\{0.64, 0.13, 0.10, 0.13\},\nonumber\\
\mathbf{G}_{2}&=&\mathrm{diag}\{0.13, 0.12, 0.62, 0.13\},\nonumber\\
\mathbf{G}_{3}&=&\mathrm{diag}\{0.11, 0.64, 0.11, 0.14\},\nonumber\\
\mathbf{G}_{4}&=&\mathrm{diag}\{0.12, 0.11, 0.12, 0.65\}.
\end{eqnarray}
These are the experimentally reconstructed coherence matrices. The data points representing these fields are plotted in Fig.~4(e) in the main text. The optical transformations used to synthesize these fields are depicted in Fig.~\ref{fig:rank4setups} and are given by:
\begin{eqnarray}
\mathbf{S}_{1}&=&\left(\begin{array}{cc}
\hat{V}_{\mathrm{PP}}(0.646,0.118)&\mathbf{0}\\
\mathbf{0}&\hat{V}_{\mathrm{ND}}(0.118)
\end{array}\right),\nonumber\\
\mathbf{S}_{2}&=&\left(\begin{array}{cc}
\hat{V}_{\mathrm{ND}}(0.118)&\mathbf{0}\\
\mathbf{0}&\hat{V}_{\mathrm{PP}}(0.646,0.118)
\end{array}\right),\nonumber\\
\mathbf{S}_{3}&=&\left(\begin{array}{cc}
\hat{V}_{\mathrm{HWP}}(45^{\circ})&\mathbf{0}\\
\mathbf{0}&\mathbb{I}
\end{array}\right)\cdot\mathbf{S}_{1},\nonumber\\
\mathbf{S}_{4}&=&\left(\begin{array}{cc}
\mathbb{I}&\mathbf{0}\\
\mathbf{0}&\hat{V}_{\mathrm{HWP}}(45^{\circ})
\end{array}\right)\cdot\mathbf{S}_{2}.
\label{eq: Rank-4 S=1.5 transformations}
\end{eqnarray}

\begin{figure*}[t!]
    \centering
    \includegraphics[width=17.6cm]{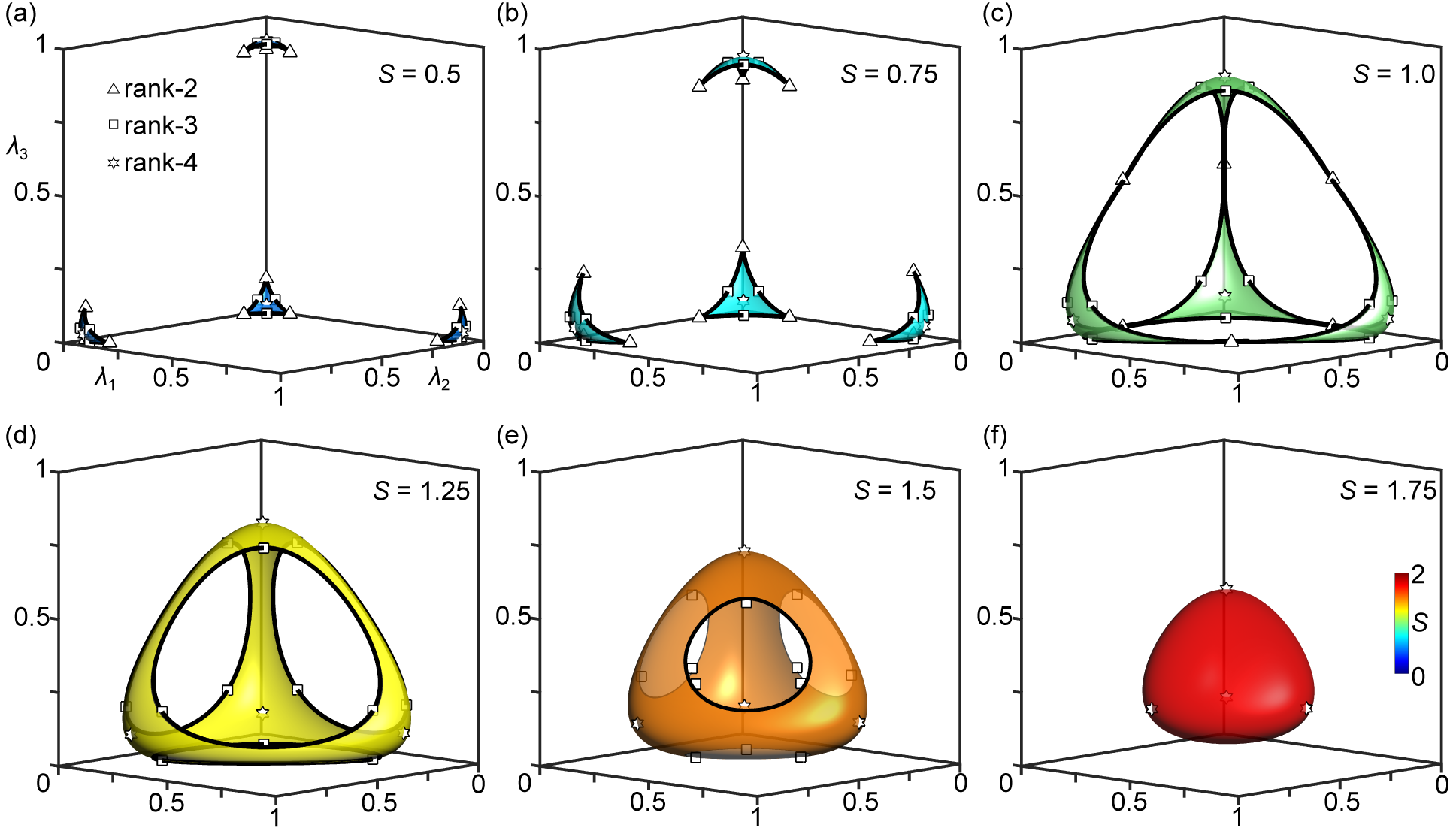}
    \caption{Theoretical coherence matrices of partially coherent optical fields of different rank represented on iso-entropy surfaces: (a) $S\!=\!0.5$, (b) $S\!=\!0.75$, (c) $S\!=\!1$, (d) $S\!=\!1.25$, (e) $S\!=\!1.5$, and (f) $S\!=\!1.75$. The inset in (a) magnifies one corner of the triangular pyramid. The white markers represent experimentally measured matrices. Rank-2 fields are represented by white triangles; rank-3 fields are represented by white squares and rank-4 fields are represented by white hexagrams.}
    \label{figS7}
\end{figure*}

\noindent\textit{Rank-4 with entropy of 1.75:} We selected the coherence matrix given by:
\begin{equation}
\mathbf{G}=\mathrm{diag}\{0.526, 0.158, 0.158, 0.158\},
\end{equation}
In our experiments, we produced optical fields represented by the following permutations of this coherence matrix:
\begin{eqnarray}
\mathbf{G}_{1}&=&\mathrm{diag}\{0.52, 0.18, 0.14, 0.16\},\nonumber\\
\mathbf{G}_{2}&=&\mathrm{diag}\{0.16, 0.16, 0.52, 0.16\},\nonumber\\
\mathbf{G}_{3}&=&\mathrm{diag}\{0.16, 0.53, 0.15, 0.17\},\nonumber\\
\mathbf{G}_{4}&=&\mathrm{diag}\{0.16, 0.16, 0.16, 0.52\}.
\end{eqnarray}
These are the experimentally reconstructed coherence matrices. The points representing these fields are plotted in Fig.~4(f) in the main text. The optical transformations used to synthesize these fields are depicted in Fig.~\ref{fig:rank4setups} and are give by:
\begin{eqnarray}
\mathbf{S}_{1}&=&\left(\begin{array}{cc}
\hat{V}_{\mathrm{PP}}(0.526,0.158)&\mathbf{0}\\
\mathbf{0}&\hat{V}_{\mathrm{ND}}(0.158)
\end{array}\right),\nonumber\\
\mathbf{S}_{2}&=&\left(\begin{array}{cc}
\hat{V}_{\mathrm{ND}}(0.158)&\mathbf{0}\\
\mathbf{0}&\hat{V}_{\mathrm{PP}}(0.526,0.158)
\end{array}\right),\nonumber\\
\mathbf{S}_{3}&=&\left(\begin{array}{cc}
\hat{V}_{\mathrm{HWP}}(45^{\circ})&\mathbf{0}\\
\mathbf{0}&\mathbb{I}
\end{array}\right)\cdot\mathbf{S}_{1},\nonumber\\
\mathbf{S}_{4}&=&\left(\begin{array}{cc}
\mathbb{I}&\mathbf{0}\\
\mathbf{0}&\hat{V}_{\mathrm{HWP}}(45^{\circ})
\end{array}\right)\cdot\mathbf{S}_{2}.
\label{eq: Rank-4 S=1.75 transformations}
\end{eqnarray}


\section{Intra-rank iso-entropy field transformations}

\textbf{For Rank-3 $S=1$:} We start with the coherence matrix $\mathbf{G}_{\mathrm{s}}\!=\!\mathrm{diag}\{\tfrac{1}{4},\tfrac{1}{4},\tfrac{1}{4},\tfrac{1}{4}\}$, convert it to $\mathbf{G}_{1}$, from $\mathbf{G}_{1}$ to $\mathrm{G}_{2}$, and then from $\mathbf{G}_{2}$ to $\mathbf{G}_{3}$:
\begin{equation}
\mathbf{G}_{\mathrm{s}}\xrightarrow{\mathbf{T}_{1}}\mathbf{G}_{1}\xrightarrow{\mathbf{T}_{2}}\mathbf{G}_{2}\xrightarrow{\mathbf{T}_{3}}\mathbf{G}_{3},
\end{equation}
where the experimentally reconstructed coherence matrices (which are equal to the theoretical target matrices within the second decimal point) are:
\begin{eqnarray}
\mathbf{G}_{1}&=&\mathrm{diag}\{0.03, 0, 0.72, 0.25\},\nonumber\\
\mathbf{G}_{2}&=&\mathrm{diag}\{0.06, 0, 0.75, 0.19\},\nonumber\\
\mathbf{G}_{3}&=&\mathrm{diag}\{0.11, 0, 0.77, 0.11\},
\end{eqnarray}
and the requisite transformations $\mathbf{T}_{1}$, $\mathbf{T}_{2}$, and $\mathbf{T}_{3}$ are given by:
\begin{eqnarray}
\mathbf{T}_{1}&=&\left(\begin{array}{cc}
\hat{V}_{\mathrm{ND}}(0.205)\cdot \hat{V}_{\mathrm{H}}&\mathbf{0}\\
\mathbf{0}&\hat{V}_{\mathrm{PP}}(1,0.597)
\end{array}\right)\nonumber\\
&=&\mathrm{diag}\{0.205, 0, 1, 0.597\},\nonumber\\
\mathbf{T}_{2}&=&\left(\begin{array}{cc}
\mathbb{I}&\mathbf{0}\\
\mathbf{0}&\hat{V}_{\mathrm{ND}}(0.835)\cdot \hat{V}_{\mathrm{PP}}(0.869,0.781)\end{array}\right)\nonumber\\
&=&\mathrm{diag}\{1, 1, 0.726, 0.602\},\nonumber\\
\mathbf{T}_{3}&=&\left(\begin{array}{cc}
\mathbb{I}&\mathbf{0}\\
\mathbf{0}&\hat{V}_{\mathrm{ND}}(0.835)\cdot \hat{V}_{\mathrm{PP}}(0.879,0.681)
\end{array}\right)\nonumber\\
&=&\mathrm{diag}\{1, 1, 0.734, 0.568\}.
\end{eqnarray}

\vspace{5pt}

\textbf{For Rank-4 $S=1.5$:}
We start with the coherence matrix $\mathbf{G}_{\mathrm{s}}\!=\!\mathrm{diag}\{\tfrac{1}{4},\tfrac{1}{4},\tfrac{1}{4},\tfrac{1}{4}\}$, convert it to $\mathbf{G}_{4}$, from $\mathbf{G}_{4}$ to $\mathbf{G}_{5}$, and then from $\mathbf{G}_{5}$ to $\mathbf{G}_{6}$:
\begin{equation}
\mathbf{G}_{\mathrm{s}}\xrightarrow{\mathbf{T}_{4}}\mathbf{G}_{4}\xrightarrow{\mathbf{T}_{5}}\mathbf{G}_{5}\xrightarrow{\mathbf{T}_{6}}\mathbf{G}_{6},
\end{equation}
where the theoretical target coherence matrices are:
\begin{eqnarray}
\mathbf{G}_{4}&=&\mathrm{diag}\{0.12, 0.12, 0.65, 0.12\},\nonumber\\
\mathbf{G}_{5}&=&\mathrm{diag}\{0.08, 0.08, 0.61, 0.23\},\nonumber\\
\mathbf{G}_{6}&=&\mathrm{diag}\{0.06, 0.06, 0.53, 0.35\},
\end{eqnarray}
and the experimental coherence matrices are:
\begin{eqnarray}
\mathbf{G}_{4}&=&\mathrm{diag}\{0.12, 0.13, 0.64, 0.11\},\nonumber\\
\mathbf{G}_{5}&=&\mathrm{diag}\{0.07, 0.09, 0.61, 0.22\},\nonumber\\
\mathbf{G}_{6}&=&\mathrm{diag}\{0.05, 0.08, 0.53, 0.35\},
\end{eqnarray}
The requisite transformations $\mathbf{T}_{4}$, $\mathbf{T}_{5}$, and $\mathbf{T}_{6}$ are given by:
\begin{eqnarray}
\mathbf{T}_{4}&=&\left(\begin{array}{cc}
\hat{V}_{\mathrm{ND}}(0.428)&\mathbf{0}\\
\mathbf{0}&\hat{V}_{\mathrm{PP}}(1,0.428)
\end{array}\right)\nonumber\\
&=&\mathrm{diag}\{0.428, 0.428, 1, 0.428\},\nonumber\\
\mathbf{T}_{5}&=&\left(\begin{array}{cc}
\hat{V}_{\mathrm{ND}}(0.595)&\mathbf{0}\\
\mathbf{0}&\hat{V}_{\mathrm{PP}}(0.704,1)
\end{array}\right)\nonumber\\
&=&\mathrm{diag}\{0.595, 0.595, 0.704, 1\},\nonumber\\
\mathbf{T}_{6}&=&\left(\begin{array}{cc}
\hat{V}_{\mathrm{ND}}(0.701)&\mathbf{0}\\
\mathbf{0}&\hat{V}_{\mathrm{PP}}(0.755,1)
\end{array}\right)\nonumber\\
&=&\mathrm{diag}\{0.701, 0.701, 0.755, 1\}.
\end{eqnarray}


\begin{figure}[h!]
\includegraphics[width=8.6cm]{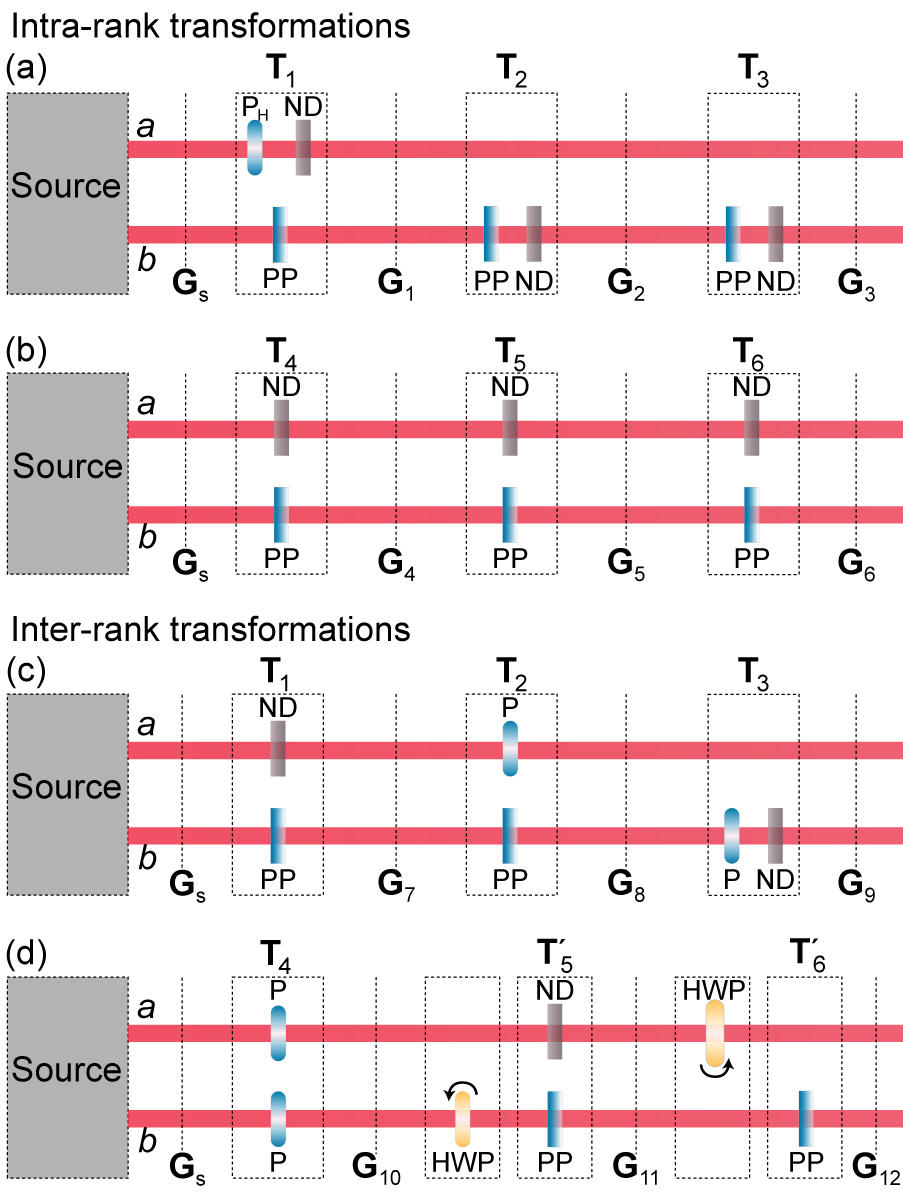}
\caption{(a) Steering intra-rank (rank-3) iso-entropy fields at $S\!=\!1$ via the transformations $\mathbf{T}_{1}$ -- $\mathbf{T}_{3}$, corresponding to Fig.~5(b,d) in the main text. (b) Steering intra-rank (rank-4) iso-entropy fields at $S\!=\!1.5$ via the transformations $\mathbf{T}_{4}$ -- $\mathbf{T}_{6}$, corresponding to Fig.~5(c,e) in the main text. (c,d) Steering inter-rank iso-entropy fields at $S\!=\!1$ via the transformations $\mathbf{T}_{1}$ -- $\mathbf{T}_{6}$, corresponding to Fig.~6 in the main text.}
\label{figS5}
\end{figure}

\begin{figure}[h!]
    \includegraphics[width=8.6cm]{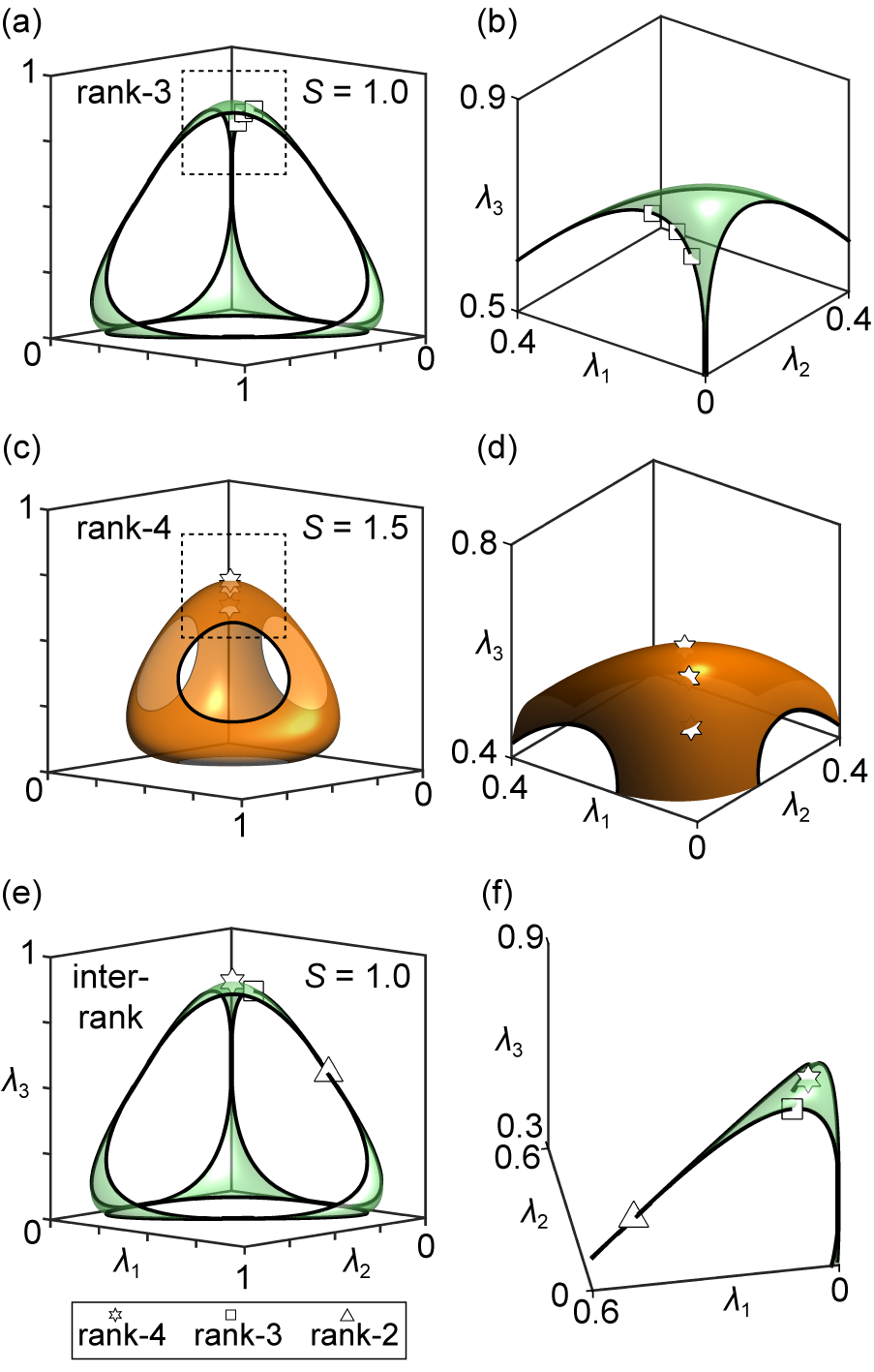}
    \caption{Plots for the theoretical coherence matrices associated with the intra-rank and inter-rank iso-entropy trajectories, corresponding to Fig.~5(b-e) and Fig.~6(a-c) in the main text.}
    \label{figS8}
\end{figure}

\section{Inter-rank iso-entropy field transformations}

Three matrices, one from each rank of 2, 3, and 4 constitute an inter-rank trajectory shown in Fig. 6 in the main text. The theoretical matrices are
\begin{eqnarray}
\mathbf{G}_{4}&=&\mathrm{diag}\{0.063, 0.063, 0.811, 0.063\},\nonumber\\
\mathbf{G}_{3}&=&\mathrm{diag}\{0.114, 0, 0.773, 0.114\},\nonumber\\
\mathbf{G}_{2}&=&\mathrm{diag}\{0.5, 0, 0.5, 0\}.
\end{eqnarray}
The required non-unitary transformations are given by:
\begin{eqnarray}
\mathbf{T}_{1}&=&\left(\begin{array}{cc}
\hat{V}_{\mathrm{ND}}(0.279)&\mathbf{0}\\
\mathbf{0}&\hat{V}_{\mathrm{PP}}(1,0.279)
\end{array}\right)\nonumber\\
&=&\mathrm{diag}\{0.279, 0.279, 1, 0.279\},\nonumber\\
\mathbf{T}_{2}&=&\left(\begin{array}{cc}
\hat{V}_{\mathrm{H}}&\mathbf{0}\\
\mathbf{0}&\hat{V}_{\mathrm{PP}}(0.728,1)
\end{array}\right)\nonumber\\
&=&\mathrm{diag}\{1, 0, 0.728, 1\},\nonumber\\
\mathbf{T}_{3}&=&\left(\begin{array}{cc}
\mathbb{I}&\mathbf{0}\\
\mathbf{0}&\hat{V}_{\mathrm{ND}}(0.383) \cdot \hat{V}_{\mathrm{H}}\end{array}\right)\nonumber\\
&=&\mathrm{diag}\{1, 1, 0.383, 0\}.
\label{Eq: T9}
\end{eqnarray}
The experimental coherence matrices are given by:
\begin{eqnarray}
\mathbf{G}_{4}&=&\mathrm{diag}\{0.06, 0.07, 0.81, 0.06\},\nonumber\\
\mathbf{G}_{3}&=&\mathrm{diag}\{0.11, 0, 0.76, 0.12\},\nonumber\\
\mathbf{G}_{2}&=&\mathrm{diag}\{0.5, 0, 0.5, 0\}.
\end{eqnarray}

For the reverse direction entropy:
\begin{eqnarray}
\mathbf{T}_{4}&=&\left(\begin{array}{cc}
\hat{V}_{\mathrm{H}}&\mathbf{0}\\
\mathbf{0}&\hat{V}_{\mathrm{H}}
\end{array}\right)\nonumber\\
&=&\mathrm{diag}\{1, 0, 1, 0\}.
\label{Eq: T10}
\end{eqnarray}
As mentioned in the main text, converting $\mathbf{G}_{2}$ to $\mathbf{G}_{3}$ and $\mathbf{G}_{3}$ to $\mathbf{G}_{4}$ require randomizing operations to invoke a third and fourth non-zero eigenvalue, respectively. By introducing a deterministic filtering operation after each randomizing one (implemented via a rotating HWP), the target entropy is restored. The following matrices ($\mathbf{T}_{5}'$ and $\mathbf{T}_{6}'$) represent the deterministic portion of these operations [see Fig.~\ref{figS5}(c,d)]:
\begin{eqnarray}
\mathbf{T}_{5}'&=&\left(\begin{array}{cc}
\hat{V}_{\mathrm{ND}}\left(0.271\right)&\mathbf{0}\\
\mathbf{0}&\hat{V}_{\mathrm{PP}}\left(1, 0.383\right)
\end{array}\right)\nonumber\\
&=&\mathrm{diag}\{0.271, 0.271, 1, 0.383\},\nonumber\\
\mathbf{T}_{6}'&=&\left(\begin{array}{cc}
\mathbb{I}&\mathbf{0}\\
\mathbf{0}&\hat{V}_{\mathrm{PP}}\left(0.972, 0.707\right)
\end{array}\right)\nonumber\\
&=&\mathrm{diag}\{1, 1, 0.972, 0.707\}
\label{Eq: T12}
\end{eqnarray}
The experimental coherence matrices are given by:
\begin{eqnarray}
\mathbf{G}_{2}&=&\mathrm{diag}\{0.50, 0.02, 0.48, 0\},\nonumber\\
\mathbf{G}_{3}&=&\mathrm{diag}\{0.11, 0.01, 0.77, 0.11\},\nonumber\\
\mathbf{G}_{4}&=&\mathrm{diag}\{0.06, 0.08, 0.80, 0.07\},
\end{eqnarray}
which are plotted in Fig.~\ref{figS8}(e,f). Note that in our experiments $\mathbf{T}_{6}'$ includes a ND filter in \textit{a} because the partial polarizer in \textit{b} resulted in power loss for both the H and V polarization modes. Partial polarizers here are constructed from a cascade of obliquely placed glass microscope cover-slips separated by thin air layers.